\documentclass[letterpaper,twocolumn,10pt]{article}
\usepackage{usenix2019_v3}

\usepackage{booktabs}
\usepackage{algorithm}
\usepackage{algorithmic}
\usepackage{xspace}
\usepackage{xcolor}
\usepackage{subcaption}
\usepackage{amsmath,amssymb,amsfonts}
\usepackage{bm}

\usepackage{graphicx}

\usepackage{bbm}

\usepackage{esvect}

\newcommand{\set}[1]{#1}

\newcommand{\prob}{\mathrm{Pr}}
\newcommand{\loss}{\mathit{l}}
\newcommand{\Loss}{\mathrm{L}}

\newcommand{\expected}[2]{\underset{#2}{\mathbb{E}}[#1]}
\newcommand{\paragraphb}[1]{\vspace{0.03in} \noindent{\bf #1} }

\newcommand{\paragraphbe}[1]{\vspace{0.03in} \noindent{\bf \em #1} }
\newcommand{\MCOMMENT}[2][.6\linewidth]{%
  \leavevmode\hfill\makebox[#1][l]{//~#2}}

\newcommand{\code}[1]{\texttt{#1}}
\graphicspath{{./images/}}

\begin{document}

\title{Blind Adversarial Network Perturbations}
\author{
{\rm Milad Nasr}\\
\and
{\rm Alireza Bahramali}\\
College of Information and Computer Sciences\\
University of Massachusetts Amherst \\
\{milad, abahramali, amir\}@cs.umass.edu
\and
{\rm Amir Houmansadr}\\
} 

\date{}
\maketitle

\begin{abstract}

Deep Neural Networks (DNNs) are commonly used for various traffic analysis problems, such as website fingerprinting and flow correlation, as they outperform traditional (e.g., statistical) techniques by large margins.
However, deep neural networks are known to be vulnerable to adversarial examples:
adversarial  inputs to the model that get labeled incorrectly by the model due to  small adversarial perturbations. 
In this paper, for the first time, we show that an adversary can defeat  DNN-based traffic analysis techniques by applying  \emph{adversarial perturbations} on the patterns of \emph{live}  network traffic.

Applying adversarial perturbations (examples) on traffic analysis classifiers faces two major challenges.
First, the perturbing party (i.e., the adversary)
should be able to apply the adversarial network  perturbations on \emph{live} traffic, with no need to buffering  traffic or having some prior knowledge about upcoming network packets. 
We design a systematic approach to creating adversarial perturbations that are  independent of their target network traffic, and therefore can be applied in real-time on live traffic. We therefore call such adversarial perturbations \emph{blind}.

Second, unlike image classification applications, perturbing traffic features  is not straight-forward as this needs to be done while preserving the correctness of dependent traffic features.   
We address this challenge by introducing  remapping functions that we use to  enforce different network constraints while creating blind adversarial perturbations.

Our blind adversarial perturbations algorithm is \emph{generic} and can be applied on various types of traffic classifiers. We demonstrate this for state-of-the-art website fingerprinting and flow correlation techniques, the two most-studied types of traffic analysis.
We also show that our blind adversarial perturbations are even \emph{transferable} between different models and architectures, so they can be applied by blackbox adversaries. 
Finally, we show that existing  countermeasures perform poorly against  blind adversarial perturbations, therefore, we introduce a tailored countermeasure. 

\end{abstract}

\section{Introduction}


Traffic analysis is the art of inferring sensitive information from the \emph{patterns} of network  traffic (as opposed to packet contents), in particular,   packet timings and  sizes.
Traffic analysis is  useful in scenarios where network traffic is encrypted, since encryption does not significantly modify traffic patterns. In particular, previous work has studied traffic analysis algorithms that  either compromise the privacy of  encrypted traffic (e.g., by linking  anonymous communications~\cite{sirinam2018deep,deepcorr}) or   enhance its security by fingerprinting malicious, obfuscated connections (e.g.,  stepping stone attacks~\cite{Yoda2000FindingAC,RAINBOW}).

Recent  advances in traffic analysis  leverage deep neural networks (DNNs) to design classifiers that are significantly (in some cases, orders of magnitude) more efficient and more reliable than traditional traffic analysis techniques. In particular, the recent website fingerprinting work of Deep Fingerprinting~\cite{sirinam2018deep}  outperforms all prior fingerprinting techniques in classifying webpages, 
and the DeepCorr~\cite{deepcorr} flow correlation technique is able to link anonymized traffic flows with accuracies  two orders magnitude superior to prior flow correlation techniques.
Given the increasing use of DNNs in traffic analysis applications, we ask ourselves the following question: \emph{can DNN-based traffic analysis techniques get defeated through    adversarially perturbing  \textemdash live\textemdash  traffic patterns?}

Note that  adversarial perturbations is an active area of research  in various  image processing applications~\cite{intriguing, DBLP:journals/corr/GoodfellowSS14, Kurakin2016AdversarialEI, dong2018boosting, DBLP:conf/cvpr/Moosavi-Dezfooli16, EAD, moosavi2017universal, he2018decision} (referred to as \emph{adversarial examples}).
However, applying adversarial perturbations on network traffic is not trivial, as it faces two major challenges.
First, the perturbing entity, i.e., the  adversary,\footnote{In our context, the adversary is not necessarily a malicious party; it is the  entity who aims to  defeat the underlying DNN traffic classifiers.}
should be able to apply his adversarial   perturbations on \emph{live} network traffic, without buffering the target traffic or knowing the patterns of upcoming network packets.
This is because in most traffic analysis applications, as will be introduced, the  adversary can \emph{not} influence the generation of target traffic, but he can  only intercept the packets  of the target traffic and perturb them on the fly.
In this paper, we are the \emph{first} to design techniques  that adversarially perturb live network traffic to defeat  DNN-based traffic classifiers; we call our approach \emph{blind adversarial perturbations}. 
Our technique  applies  adversarial  perturbations on live packets as they appear on the wire.
The key idea of our adversarial perturbations algorithm is that it generates ``blind'' perturbations that are  \emph{independent of the target inputs} by solving specific optimization problems that will be explained.
We design adversarial perturbation mechanisms for  the key features commonly used   in traffic analysis applications:
our adversarial perturbations include changing the
 timings and sizes of packets, as well as inserting dummy network packets.

The second challenge to applying adversarial perturbations on traffic analysis applications is that, any perturbation mechanism on network traffic  should preserve various constraints of traffic patterns, e.g.,  the dependencies between different traffic features, the statistical distribution of timings/sizes expected from the underlying protocol, etc.
This is unlike traditional adversarial example studies (in the context of image processing)
 that modify image pixel values \emph{individually}.
 Therefore,    one can not simply borrow techniques from traditional adversarial examples.
We consequently design various \emph{remapping functions} and \emph{regularizers}, that we incorporate into our optimization problem to enforce such network constraints.
As will be shown, in most  scenarios the constraints  are \emph{not differentiable},  and therefore we carefully craft   \emph{custom gradient functions} to approximate their gradients.



\paragraphb{Evaluations:}  Our blind adversarial perturbations algorithm is \emph{generic} and can be applied to various types of traffic classifiers.
We demonstrate this by applying our techniques on state-of-the-art website fingerprinting~\cite{sirinam2018deep, var-cnn} and flow correlation\cite{deepcorr} techniques, the two most-studied types of traffic analysis.
Our evaluations show that our adversarial perturbations can effectively defeat DNN-based traffic analysis techniques through  small, live adversarial perturbations.
For instance, our perturbations can reduce the  accuracy of state-of-the-art website fingerprinting~\cite{var-cnn,sirinam2018deep} works by 90$\%$ by only adding 10$\%$ bandwidth overhead. 
Also, our adversarial perturbations  can reduce the true positive rate  of state-of-the-art flow correlation techniques~\cite{deepcorr} from 0.9 to 0.2 by applying  tiny delays with a 50ms jitter standard deviation.

We also show that our blind adversarial perturbations are  \emph{transferable} between different models and architectures, which signifies their practical importance as they can be implemented by blackbox adversaries.

\paragraphb{Countermeasures:}
We conclude by  studying various countermeasures against our adversarial perturbations.
We start by leveraging existing defenses against adversarial examples from the image classification literature and adapting them to the traffic analysis scenario. We show that such adapted defenses are \emph{not effective} against our  network adversarial perturbations as they do not take into account the specific constraints of traffic features.
Motivated by this, we design a tailored  countermeasure  for our network adversarial perturbations, which we demonstrate  to be more effective than the adapted defenses. 
The key idea of our countermeasure is  performing  adversarial training, and   using our attack as a regularizer to train robust traffic analysis models.

%
%


\section{Preliminaries}

\subsection{Problem Statement}

\emph{Traffic analysis} is to infer sensitive information from the patterns of network traffic, i.e., packet timings and sizes. Therefore, many works have investigated the use of traffic analysis in various scenarios where traffic contents are encrypted.
In particular, traffic analysis has been used to compromise anonymity in anonymous communications systems through various types of attacks, specifically, website fingerprinting~\cite{sirinam2018deep, var-cnn, rimmer2017automated, 184463, panchenko2011website, cai2012touching, wang2013improved, panchenko2016website, hayes2016k, wang2016realistically},  and flow correlation~\cite{deepcorr, CompressiveTA, RAINBOW, chothia2011statistical, zhu2004flow, levine2004timing, shmatikov2006timing, houmansadr2014non, nasr2017compressive, raptor}.    Traffic analysis has also been used to trace back cybercriminals who obfuscate their identifies through stepping stone relays~\cite{RAINBOW,houmansadr2014non,deepcorr,Yoda2000FindingAC,Zhangstep}.


\paragraphb{Our problem: Defeating DNN-based  traffic analysis algorithms}
The state-of-the-art traffic analysis techniques use deep neural networks (DNN) to offer much higher performances than prior techniques. For instance, DeepCorr~\cite{deepcorr} provides a flow correlation accuracy of $96\%$ compared to $4\%$ of statistical-based systems like RAPTOR~\cite{raptor} (in a given setting).
Also, Var-CNN~\cite{var-cnn} leverages deep learning techniques to perform a website fingerprinting attack which achieves $98\%$ accuracy in a closed-world setting.
However, deep learning models are infamous for being susceptible to various adversarial attacks where the adversary adds  small perturbations to the inputs to mislead  the deep learning model. Such techniques are known as \emph{adversarial examples} in the context of image processing, but have not been investigated in the traffic analysis domain.
In this work, we study the possibility of defeating DNN-based  traffic analysis techniques  through adversarial perturbations.


In our setting, some \textbf{traffic analysis parties} use  DNN-based traffic analysis techniques for various purposes, such as breaking Tor's anonymity or detecting cybercriminals. On the other hand, the \textbf{traffic analysis adversary(ies)}  aim at interfering with  the traffic analysis process through adversarially perturbing traffic patterns of the connections they intercept.
To do so, the traffic analysis adversary(ies) perturb the traffic patterns of the intercepted flows to reduce the accuracy of the DNN-based classifiers  used by the traffic analysis parties.
To further clarify the distinction between the players, in the flow correlation setting, the traffic analysis ``party'' can be a malicious ISP who aims at deanonymizing Tor users by analyzing their Tor connections; however, the traffic analysis ``adversary'' can be some (benign) Tor relays who perturb traffic patterns of their connections to defeat potential traffic analysis attacks.

\paragraphb{Challenges:} Note that our problem resembles the setting of adversarial examples for image classification. However, applying adversarial perturbations on network traffic presents two major challenges.
First, the adversaries should be able to apply  adversarial perturbations  on \emph{live} network connections where the patterns of upcoming network packets are \emph{unknown} to the adversaries. This is because in traffic analysis applications, the adversary is not in charge of generating traffic patterns. For instance, in  the flow correlation scenario, the traffic analysis adversary is a benign Tor relay  who intercepts and (slightly) perturbs \emph{the traffic generated by  Tor users}.
The second challenge to applying network adversarial perturbations is that they should preserve the various constraints of network traffic, e.g., the dependencies of different traffic features.



\begin{figure}[!t]
    \centering
    \includegraphics[width = 1\linewidth]{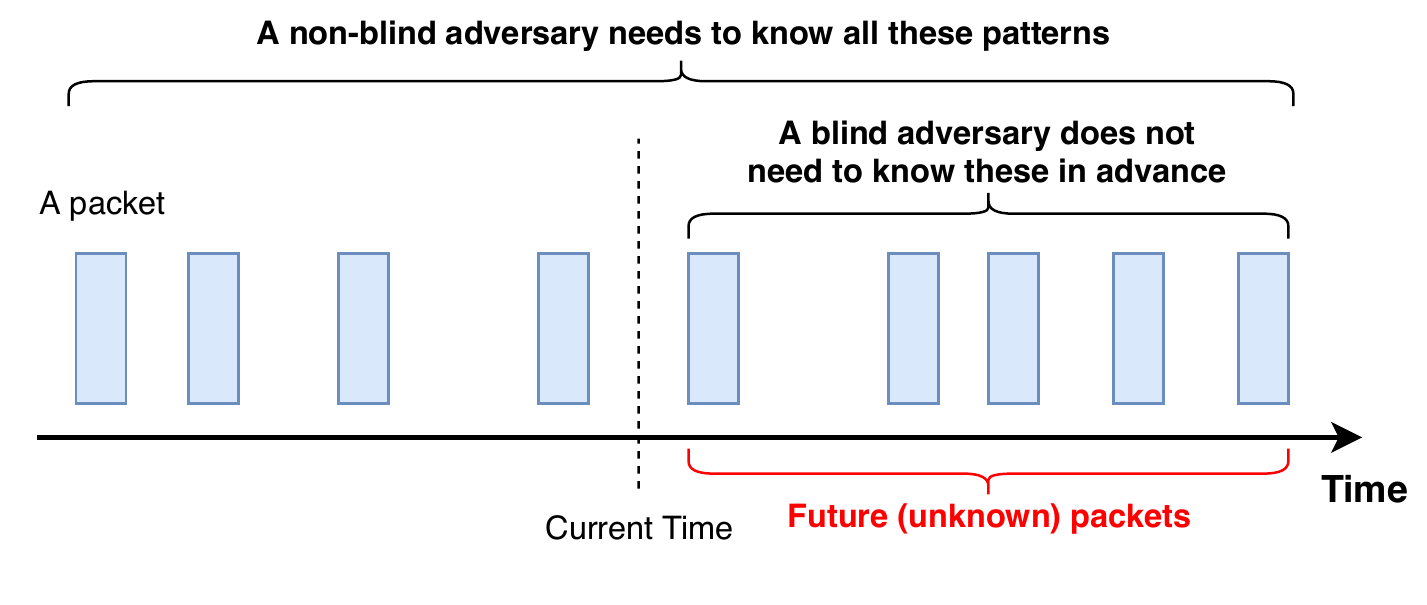}
    \caption{Unlike traditional adversarial perturbation techniques, our \emph{blind} perturbation approach does not need to know the patterns of its target connection.}
    \label{fig:blind}
\end{figure}

\paragraphb{Sketch of our approach:}
In this work, we design \emph{blind adversarial perturbations}, a set of techniques to perform adversarial network perturbations that overcome the two mentioned challenges.
To address the first challenge (applying on live traffic), we design \emph{blind} perturbation vectors that are independent of their target traffic, therefore, they can be applied on any (unknown) network flows.
Figure~\ref{fig:blind} shows what is needed by our blind perturbations adversary compared to traditional (non-blind) perturbation techniques.
We generate such blind  adversarial perturbations by solving a specific  optimization problem.
We address the second challenge (enforcing network constraints) by using various \emph{remapping functions} and \emph{regularizers} that adjust perturbed traffic features to follow the required constraints.

\subsection{Threat Model}\label{sec:threat}

\paragraphb{Adversary's knowledge of the inputs.} 
We assume the adversary has \emph{no prior knowledge} about the patterns of upcoming network packets of the target connections to be perturbed. 
Therefore,  the goal of the adversary is to craft
 input-agnostic perturbation vectors that once applied to arbitrary network flows, such flows are misclassified by the target 
   DNN-based traffic analysis algorithms.

\paragraphb{Adversary's Knowledge of the model.}
We start with an adversary who has a white-box access to the target model to be defeated, i.e., he knows the target DNN model's architecture and parameters (Section~\ref{sec:method}).
Then, in Section~\ref{sec:transfer} we extend our attack to  a blackbox setting where the adversary has no knowledge of the target model's architecture or parameters, by leveraging the transferability of our  technique.


\paragraphb{Adversary's Knowledge of the training data.}
 We  assume the adversary knows a  set of samples from the same distribution as the training dataset of the target model. For example, in the website fingerprinting application the adversary can browse the target websites to be misclassified to obtain such training samples.

\paragraphb{Attack's target.}
We consider four types of attacks, i.e.,  \code{ST}-\code{DT}, \code{ST}-\code{DU}, \code{SU}-\code{DT}, and \code{SU}-\code{DU}, based on the adversary's source and destination targets as defined below:

\paragraphbe{a) Destination-targeted/untargeted (\code{DT}/\code{DU}):} We call the attack \emph{destination-targeted} (\code{DT})  if the goal of the adversary is to make the  model  misclassify arbitrary inputs into a specific, target output class. On the other hand, we call the attack \emph{destination-untargeted} (\code{DU})  if the goal is to misclassify inputs into arbitrary (incorrect) output classes.

\paragraphbe{b) Source-targeted/untargeted (\code{ST}/\code{SU}): } A \emph{source-targeted} (\code{ST}) adversary  is one whose goal is to have inputs from a specific input class  misclassified by  the traffic analysis model. By contrast, a \emph{source-untargeted} (\code{SU}) adversary is one who aims at  causing arbitrary inputs classes to get misclassified.

\section{Background}

\subsection{Deep learning}
A deep neural network  (DNN) consists of a series of linear and nonlinear functions,  known as layers. Each layer has a weight matrix $w_i$ and an activation function. For a given input $\bm{x}$, we  denote the output of  a DNN model by:
\begin{align*}
  f(\bm{x})= f_n^{w_n}(f_{n-1}^{w_{n-1}}( \cdots  f_1^{w_1}(x1\bm{x})))\cdots)
\end{align*}
where $f_i^{w_i}$ is the $i-$th layer of the deep neural network (note that we use bold letters to represent vectors as in $\bm{x}$).
We focus on supervised learning, where we have a set of labeled training data.  Let $\set{X}$ be a set of data points in the target $d$-dimensional space, where each dimension represents one attribute of the input data points. We assume there is an oracle $O$ which maps the data points to their labels. For the sake of simplicity, we only focus on  classification tasks.

The goal of training  is to find a classification model $f$ that  maps each point in $\set{X}$ to its correct class in the set of classes, $\set{Y}$. To obtain $f$,  one needs to define a lower-bounded, real-valued loss function $\loss(f(\bm{x}), O(\bm{x}))$ that for each data point $\bm{x}$ measures the difference between $O(\bm{x})$ and the model's prediction $f(\bm{x})$.

Therefore, the loss function for $f$ can be defined as:
\begin{align}\label{eq:closs}
   \Loss(f) &= \expected {\loss (f( \bm{x} ),y)}{(\bm{x},y) \sim \prob(\set{X}, \set{Y})}
\end{align}
and the objective of training is to find an $f$ that minimizes this loss function.
Since  $\prob(\set{X}, \set{Y})$ is not entirely available to the training entities, in practice, a set of samples from it,  called the training set $\set{D}^{train} \subset \set{X} \times \set{Y}$,
is used to train the model~\cite{vapnik2013nature}.
Therefore, instead of minimizing \eqref{eq:closs}, machine learning algorithms minimize the expected {\em empirical loss} of the model over its training set $\set{D}$:
\begin{align}\label{eq:closs_emp}
   \Loss_{\set{D}^{train}}(f) = \frac{1}{|\set{D}^{train}|}\sum\limits_{(\bm{x}, y) \in \set{D}^{train}} \loss(f(\bm{x}), y)
\end{align}
Therefore, a deep neural network $f$ is trained by tuning its weight parameters to
minimize its empirical  loss function over a
 (large) set of known input-output pairs $(x, y)$.
This is commonly performed using a variation of the gradient descent algorithm, e.g.,  back propagation~\cite{goodfellow2016deep}.



\subsection{Adversarial Examples}
An adversarial example is an adversarially crafted input that fools  a target
classifier or regression model into making incorrect classifications or predictions.
The adversary's goal is to generate adversarial examples by adding minimal perturbations to the input data attributes. Therefore, an adversarial example $\bm{x}^*$ can be crafted by solving the following  optimization problem:
\begin{align}\label{eq:erm_untargted}
  \bm{x}^* = \bm{x} + \arg \min \{ \bm{z}:O(\bm{x}+\bm{z}) \neq O(\bm{x}) \} = \bm{x} + \bm{\delta_{x}}
\end{align}
where $x$ is a non-adversarial input sample,

 $\bm{\delta_{x}}$ is the adversarial perturbation added to it, and  $O(\cdot)$ represents   the true label of its input, as defined in the previous section. The adversary's objective is to adds a minimal perturbation $\bm{\delta_{x}}$ to force the target model misclassify the input $\bm{x}$. Adversarial examples are commonly studied in image classification applications, where a constraint in finding adversarial examples is that the added noise should be imperceptible to the human eyes.

 In this paper, we will investigate the application of adversarial examples on network connections with different imperceptibility constraints.


Previous works~\cite{DBLP:journals/corr/GoodfellowSS14, goodfellow2014explaining, kurakin2016adversarial, dong2018boosting, moosavi2017universal} have suggested several ways to generate adversarial examples.
The Fast Gradient Sign Method (FGSM)~\cite{DBLP:journals/corr/GoodfellowSS14} algorithm
generates an adversarial sample by calculating the following perturbation for a given  model $f$ and a loss function $\loss$:
\begin{align}
   \bm{\delta_{x}} = \epsilon \times \text{Sign} (\nabla_{\bm{x}} \loss(f(\bm{x}),y) )
\end{align}
where $\nabla_{\bm{x}} l(f(\bm{x}),y)$ is  the model's loss gradient w.r.t.\ the input $\bm{x}$, and the $y$ is the input's label.
Therefore, the adversarial perturbation is the sign of the model's loss gradient w.r.t. the input $\bm{x}$ and label $y$.
 Also, $\epsilon$ is a coefficient controlling the amplitude of the perturbation. Therefore, the adversarial perturbation in FGSM is the sign of model's gradient. The adversary adds the perturbation to  $x$ to craft an adversarial example.
 Kurakin et al.~\cite{kurakin2016adversarial} proposed a \emph{targeted} version of FGSM, where the adversary's goal is to fool the model to classify inputs as a desired target class (as opposed to \emph{any} class in FGSM). Kurakin  et al.\ also introduced an iterative method to improve the success rate of the generated examples.
 Dong et al.~\cite{dong2018boosting} showed that using the momentum approach can improve  Kurkain et al.'s iterative method.
Also, Carlini and Wagner~\cite{carlini2017towards} designed a set of attacks that can craft adversarial examples when the adversary has various norm constraints (e.g., $l_0$, $l_1$, $l_{\infty}$).
Other variations of  adversarial examples~\cite{onepixel,  eykholt2018robust} have been introduced to craft adversarial examples that consider different sets of constraints or improve the adversary's success rate.
 Moosavi-Dezfooli et al.~\cite{moosavi2017universal} introduced universal adversarial perturbations where the adversary generates adversarial examples that are  independent of the inputs.

\subsection{Traffic Analysis Techniques}
We overview the two major classes of traffic analysis techniques, which we will use to demonstrate our network adversarial perturbations.


\paragraphb{Flow correlation:}
Flow correlation aims at linking obfuscated network flows by correlating their traffic characteristics, i.e., packet timings and sizes~\cite{CompressiveTA, RAINBOW, deepcorr, bahramali2020}. In particular, the Tor anonymity system has been the target of flow correlation attacks, where an adversary aims at linking ingress and egress segments of a Tor connection by correlating traffic characteristics.
Traditional flow correlation techniques mainly use standard statistical correlation metrics to correlate the vectors of flow timings and sizes across flows, in particular mutual information~\cite{chothia2011statistical, zhu2004flow}, Pearson correlation~\cite{levine2004timing, shmatikov2006timing}, cosine similarity~\cite{houmansadr2014non, nasr2017compressive}, and Spearman correlation~\cite{raptor}.
More recently, Nasr et al.~\cite{deepcorr} design a DNN-based approach for flow correlation, called DeepCorr. They show that DeepCorr
outperforms  statistical flow correlation techniques by large margins.


\paragraphb{Website Fingerprinting:} Website fingerprinting (WF) aims at detecting
the websites  visited over encrypted channels
like VPNs, Tor, and other proxies~\cite{sirinam2018deep, var-cnn, rimmer2017automated, 184463, panchenko2011website, cai2012touching, wang2013improved, panchenko2016website, hayes2016k, wang2016realistically}.
The attack is performed by a passive adversary who monitors the victim's encrypted network traffic, e.g., a malicious ISP or a surveillance agency.
The adversary compares the victim's observed traffic
flow against a set of prerecorded webpage traces, to identify the webpage being browsed.
 Website fingerprinting differs from flow correlation in that
the adversary only observes one end of the connection, e.g., the connection between a client and a Tor relay.
Website fingerprinting has been widely studied in the
context of Tor traffic analysis~\cite{rimmer2017automated, sirinam2018deep, var-cnn, panchenko2011website, cai2012touching, wang2013improved, panchenko2016website, hayes2016k}.

Various machine learning classifiers have been used for WF, e.g.,  using KNN~\cite{184463}, SVM~\cite{panchenko2016website}, and random forest~\cite{hayes2016k}.
However, the state-of-the-art WF algorithms  use Convolutional Neural Networks to perform website fingerprinting, i.e., Sirinam et al.~\cite{sirinam2018deep}, Rimmer et al.~\cite{rimmer2017automated}, and Bhat et al.~\cite{var-cnn}.

\paragraphb{Defenses:}
Note that our blind adversarial perturbations technique serves as a defense mechanism against traffic analysis classifiers (as it aims at fooling the underlying classifiers). 
The literature has proposed other defenses against website fingerprinting and flow correlation attacks~\cite{juarez2016toward,wang2017walkie,cherubin2017website,cai2014cs}.
Similar to our work, such defenses work by manipulating traffic features, i.e., packet timings, sizes, and directions. 

In Section~\ref{sec:comp-defense}, we compare the performance of our blind adversarial perturbations with state-of-the-art defenses, showing that our technique outperforms all of these techniques in defeating traffic analysis. 



Also, note that some recent works have considered using adversarial perturbations as a defense against traffic analysis. 
In particular,  Mockingbird~\cite{mockingbird} generates adversarial perturbations to defeat website fingerprinting, and 
Zhang et al.~\cite{zhang2019statistical} apply adversarial examples to defeat video classification using traffic analysis.
However, \emph{both of these works are non-blind}, i.e., the adversary needs to know the patterns of the target flows in advance; therefore, we consider them to be unusable in typical traffic analysis scenarios. By contrast, our blind perturbation technique modifies live network connections.


\section{Formalulating Blind Adversarial Perturbations} \label{sec:method}



In this section, we present the key formulation and algorithms for generating blind adversarial perturbations.

\subsection{The General Formulation}


 We  formulate the  blind adversarial perturbations problem as the following optimization problem:
\begin{align}\label{eq:adv_main}
   \arg \min_{\bm{\delta}}  \forall \bm{x} \in D^S : f(\bm{x}+\bm{\delta}) \neq f(\bm{x})
\end{align}
where the objective is to find
a (blind) perturbation vector, $\bm{\delta}$, such that when added to an \emph{arbitrary} input from a target input domain $D^S$, it will cause the underlying DNN model $f(.)$ to misclassify.
In a source-targeted (\code{ST}) attack (see definitions in Section~\ref{sec:threat}), $D^S$ contains inputs from a  target class to be misclassifies, whereas in a source-untargeted (\code{SU}) attack  $D^S$ will be a large set of inputs from different classes.



Note that one cannot find a closed-form solution for this optimization problem since the target model $f(.)$ is a non-convex ML model, i.e., a deep neural network.
Therefore, \eqref{eq:adv_main} can be formulated as follows to numerically solve the problem using  \emph{empirical approximation techniques}:
\begin{align}\label{eq:erm_untargted}
    &\arg \max_{\bm{\delta}}   \sum_{\bm{x} \in \mathcal{D}^{S}}  \loss(f(\bm{x}+\bm{\delta}),f(\bm{x}))
\end{align}
where $\loss$ is the target model's loss function and $\mathcal{D}^S\subset D^S$ is the  adversary's network training dataset. 

Note that prior work by 
Moosavi-Dezfooli et al.~\cite{moosavi2017universal} has
 studied the generation of  universal adversarial perturbations for  image recognition applications.
We, however, take a different direction in generating blind perturbations: in contrast to finding a perturbation vector $\bm{\delta}$ that  maximizes the loss function in \cite{moosavi2017universal},
  \emph{we aim to find  a perturbation generator model $G$}.
 This generator model $G$ will generate adversarial perturbation vectors when provided with a random \emph{trigger} parameter $z$ (we denote the corresponding adversarial perturbation as $\bm{\delta_z}=G(z)$), i.e., we are able to generate different perturbations on different random $z$'s.
 Therefore, the goal of our optimization problem is
  to optimize the parameters of the perturbation generator model $G$  (as opposed to optimizing a perturbation vector $\bm{\delta}$  in \cite{moosavi2017universal}). 
  Using a generator model increases the attack performance, as shown previously~\cite{hayes2018learning, abdoli2019universal} and validated through our experiments.
  Hence, we formulate our optimization problem as:
\begin{align}\label{eq:erm_untargted}
    & \arg \max_{G}   \expected{ \sum_{\bm{x} \in \mathcal{D}^{S   }}  \loss(f(\bm{x}+G(z)),f(\bm{x}))}{ z \sim uniform (0,1)}
\end{align}

 We can use existing optimization techniques (e.g., Adam~\cite{adam}) to  solve this problem.
 In each iteration of  training, our algorithm selects a batch from the training dataset and a random trigger $z$,  then computes the objective function. 

\subsection{Incorporating  Traffic Constraints}

Studies of adversarial examples for image recognition applications~\cite{goodfellow2014explaining, kurakin2016adversarial, dong2018boosting, DBLP:journals/corr/GoodfellowSS14, moosavi2017universal}
simply modify image pixel values \emph{individually}.
However, applying adversarial perturbations on network traffic is much more challenging due to the various \emph{constraints} of network traffic that should be preserved while applying the perturbations.
In particular,  inter-packet delays  should have non-negative values;
	 the target network  protocol may need to follow specific  packet size/timing distributions;
	  packets should not be removed from a connection; and, packet numbers should get adjusted after injecting new packets.

%

One can add other network constraints depending on the underlying network protocol. 
We  use \emph{remapping and regularization functions}  to enforce  these domain constraints while creating blind adversarial perturbations. A remapping function adjusts the perturbed traffic patterns so they comply with some domain constraints.
For example, when an adversary adds a packet to a traffic flow at  position $i$, the remapping function should shift the indices of all  consecutive packets.

We therefore reformulate our optimization problem by including the remapping function $\mathcal{M}$:
\begin{align}\label{eq:erm_untargted}
    &\arg \max_{G} \expected{  \sum_{\bm{x} \in \mathcal{D}^{S}}\loss(f(\mathcal{M}(\bm{x},G(z))),f(\bm{x}))}{z \sim uniform (0,1)}
\end{align}

 Moreover, we add a regularization term to the loss function so that the adversary can enforce additional constraints, as will be discussed. Therefore, the following is our complete optimization problem:
 \begin{align}\label{eq:erm_untargted}
    &\arg \max_{G} \expected{ ( \sum_{\bm{x} \in \mathcal{D}^{S   }} \loss (f(\mathcal{M}(\bm{x},G(z))),f(\bm{x})) )  + \mathcal{R}(G(z))}{z \sim uniform (0,1)}
\end{align}

We adjust \eqref{eq:erm_untargted} for a destination-targeted (\code{DT}) attack by replacing  $\loss(f(\mathcal{M}(\bm{x},G(z))),f(\bm{x}))$  with $-\loss(f(\mathcal{M}(\bm{x},G(z))),O_T)$, where $O_T$ is the  target output class.
Also, remind that for source-targeted attacks, $\mathcal{D}^{S}$ contains samples only from the target classes. 
%

   \subsection{Overview of the Algorithm}

   Algorithm~\ref{alg:general_adv} summarizes our approach to generate blind adversarial perturbations (Figure~\ref{fig:training} in Appendix~\ref{app:scheme} illustrates  the main components of our algorithm).
   In each iteration, Algorithm~\ref{alg:general_adv} computes the gradient of the objective function w.r.t.\  the blind perturbation for given inputs, and optimizes it by moving in the direction of the gradient. 
   The algorithm enforces domain constraints using  various remapping and regularization functions.
   We use the iterative mini-batch stochastic gradient ascent~\cite{goodfellow2016deep} technique.

\begin{algorithm}[t!]
\begin{footnotesize}
    \caption{Generating Blind Adversarial Perturbations}
   \begin{algorithmic}
       \STATE $\mathcal{D}^{S} \gets$ adversary training data
       \STATE $f \gets$ target model
       \STATE $\mathcal{L}_{f} \gets $ target model loss function
       \STATE $\mathcal{M} \gets$ domain  remapping function
       \STATE $\mathcal{R} \gets$ domain  regularizations function
       \STATE $G(z) \gets$ initialize the blind adversarial perturbation model parameters ($\theta_G$)
       \STATE $T \gets $ epochs
       \STATE DT $ \gets$  the destination target class or false o.w.
       \STATE ST $ \gets$  the source target classes or false o.w.
       \FOR{epoch $t \in \{ 1 \cdots T \}$  }
           \FORALL{mini-batch $b_i$ in $\mathcal{D}^{S}$}
               \IF{ST }
               \STATE $b_i \gets$  select instances only with the ST class label
               \ENDIF
               \STATE $z \sim$ Uniform
               \IF{DT}
               \STATE $J = -(\frac{1}{|b_i|} \sum_{\bm{x}  \in b_i} \loss (f(\mathcal{M}(\bm{x},G(z))),f(\bm{x})) ) + \mathcal{R}(G(z)) $
               \ELSE
               \STATE $J = (\frac{1}{|b_i|} \sum_{\bm{x}  \in b_i, } \loss (f(\mathcal{M}(\bm{x},G(z))),\text{DT}) ) + \mathcal{R}(G(z))$
               \ENDIF
               \STATE  Update $G$ to minimize $J$
           \ENDFOR
       \ENDFOR
       \RETURN $G$
    \end{algorithmic}
    \label{alg:general_adv}
\end{footnotesize}
\end{algorithm}

\section{Perturbation Techniques}

The traffic analysis literature uses three main features for building traffic analysis classifiers:
 1) \emph{packet timings}~\cite{deepcorr, var-cnn}, 2)  \emph{packet sizes}~\cite{deepcorr}, and  3) \emph{packet directions}~\cite{var-cnn, sirinam2018deep, rimmer2017automated, 184463}.
 Our blind adversarial perturbation technique leverages these features to adversarially perturb traffic.
 These features can be modified either by delaying packets, resizing packets, or injecting new (dummy) packets (dropping packets is not an option as it will break the underlying applications).
We describe how we perform such perturbations.

\subsection{Manipulating Existing Packets}

The adversary can modify the timings and sizes (but not the directions) of existing packets of a target network connection.
We present a network connection as a vector of features:   $\bm{F} = [f_1,f_2,\cdots, f_n]$, where $f_i$ can represent  the size, timing, direction, or a combination of these features for the $i$th packet.
The adversary designs a blind adversarial perturbation model $G$, as introduced in Section~\ref{sec:method}, such that it outputs a perturbation  vector $G(z)=[g_1,g_2,\cdots,g_n]$ with the same size as $\bm{F}$.
The adversary adds $G(z)$ to the original traffic patterns as packets arrive, so $\bm{F}^p=\bm{F}+G(z) = [f_1+g_1,f_2+g_2,\cdots,f_1+g_n]$ is the patterns of the perturbed connection.
The main challenge is that the perturbed traffic features, $\bm{F}^p$, should not violate the domain constraints of the target network application.


\paragraphb{Perturbing timings:}
We first introduce how the timing features can be perturbed. We use inter-packet delays (IPDs) to represent the timing information of packets.
An important constraint on the timing features is that \emph{the adversary should not introduce excessive delays on the packets} as excessive delays will either interfere with the underlying application (e.g., Tor relays are not willing to introduce large  latencies) or give away the adversary.
We control the amount of delay added by the adversary  by using a  \emph{remapping function} $\mathcal{M}^T$ as follows:
\scalebox{.95}{\parbox{\linewidth}{
\begin{align}
	&\mathcal{M}^T (\bm{x},G(z),\mu, \sigma) =   \bm{x} + \nonumber \\
	&\frac{G(z) - \max(\overline{G(z)}-\mu,0) - \min(\overline{G(z)}+\mu,0)}{\text{std}(G(z))}  \min(\text{std}(G(z)),\sigma)
\end{align}}}
\noindent where $\overline{G(z)}$ is the mean of perturbation $G(z)$, and $\mu$ and  $\sigma$ are  the maximum allowed average and standard deviation of the delays, respectively. Using this remapping function, we can govern the amount of latency added to the packets.

A second  constraint on timing features is that \emph{the perturbed timings should follow the  statistical distributions expected from the target protocol}.
Towards this, we leverage a \emph{regularizer} $\mathcal{R}$ to enforce the desired statistical behavior on the blind perturbations.
Our regularizer enforces a Laplacian distribution for network jitters, as suggested by prior work~\cite{CompressiveTA}, but it can enforce arbitrary distributions.
To do this, we use a  generative adversarial network (GAN)~\cite{NIPS2014_5423}:  we design a discriminator model $D(G(x))$  which tries to distinguish the generated perturbations from a Laplace distribution. Then, we use this discriminator as our regularizer function to make the distribution of the crafted perturbations similar to a Laplace distribution. We simultaneously train the blind perturbation model and the discriminator model. This is summarized in Algorithm~\ref{alg:laplace}.

\begin{algorithm}[t!]
\begin{footnotesize}
    \caption{GAN-based timing regularizer}
   \begin{algorithmic}
       \STATE $\mathcal{D}^{S} \gets$ adversary training data
       \STATE $f \gets$ target model
       \STATE $G \gets$ blind adversarial perturbation model
       \STATE $D \gets$ discrimination model
       \STATE $\mu,b \gets$ target desired Laplace distribution parameters
       \FOR{$t \in \{1,2,\cdots,T\}$}
	   \STATE $z' \sim \text{Lapace}(\mu,b)$
	   \STATE $z \sim \text{Uniform}()$
	   \STATE train $D$ on $G(z)$ with label 1 and $z'$ with label 0
	   \STATE train $G$ on $\mathcal{D}^{S}$ using regularizer $D$
       \ENDFOR
       \RETURN $z$
    \end{algorithmic}
    \label{alg:laplace}
  \end{footnotesize}
\end{algorithm}

\paragraphb{Perturbing sizes:}
An adversary can perturb packet sizes by  increasing  packets sizes (through appending dummy bits).
However, modified packet sizes should not violate the expected  maximum packet size  of the underlying protocol as well as the expected statistical distribution of the sizes. For instance, Tor packets are expected to have certain packet sizes.


 \begin{algorithm}[t!]
\begin{footnotesize}
    \caption{Size remapping function}
   \begin{algorithmic}
       \STATE $a \gets G(z)$
       \STATE $\bm{x} \gets $ input
       \STATE $N \gets$ maximum sum of added sizes
       \STATE $n \gets$ maximum added size to each packet
       \STATE $s \gets$ cell sizes

       \FOR{ $i$ in argsort(-a)}
       \IF {$N \leq$ 0 }
       \STATE break
       \ENDIF
       \STATE $\delta =   \lfloor \min(s\frac{a[i]}{s},n,N) \rfloor $
       \STATE $N = N - \delta$
       \STATE $\bm{x}[i] = \bm{x}[i] + \delta$
       \ENDFOR
       \RETURN $\bm{x}$
    \end{algorithmic}
    \label{alg:size_g}
\end{footnotesize}
\end{algorithm}

We use the remapping function $\mathcal{M}^S$, as shown in Algorithm~\ref{alg:size_g},  to adjust the amplitude of size modifications as well as to enforce the desired statistical  distributions.
The input to Algorithm~\ref{alg:size_g} is the  blind adversarial perturbation ($G(z)$), the desired maximum bytes of added traffic  ($N$), the desired maximum added bytes to a single packet ($n$), and the expected packet size distribution of the underlying network protocol ($s$) (if the network protocol does not have any specific size constraints, then $s=1$).
Algorithm~\ref{alg:size_g} starts by selecting the highest values from the output of the adversarial perturbations and adds them to the traffic flows up to $N$ bytes. Since Algorithm~\ref{alg:size_g} is not differentiable, we cannot simply use  Algorithm~\ref{alg:general_adv}. Instead, we define a custom gradient function for Algorithm~\ref{alg:size_g} which allows us to train the blind adversarial perturbation model. Given the gradient of the target model's loss w.r.t.\ the output of Algorithm~\ref{alg:size_g} (i.e., $\nabla_{\bm{x}} {\mathcal{M}^S(\bm{x},G(z))}$), we modify  the perturbation model's gradient as:
\begin{align}
	\nabla_{G(z)} = \sum_{\bm{x} \in b_i}\nabla_{\bm{x}} {\mathcal{M}^S (\bm{x},G(z))}
\end{align}
where $b_i$ is the selected training batch. We do not need  regularization for packet sizes.

\subsection{Injecting Adversarial Packets}

In addition to perturbing the features of existing packets, the adversary can also inject packets with specific sizes and at specific times into the target connection to be perturbed.
The goal of our adversary is to identify the most adversarial timing and sizes for injected packets.
We design a  remapping function $\mathcal{M}^I$ (Algorithm~\ref{alg:insert}) that obtains the  ordering of injected packets as well as their feature values.
Similar to the previous attack, Algorithm~\ref{alg:insert} is not differentiable and we cannot simply use it for Algorithm~\ref{alg:general_adv}. Instead, we use a custom gradient function for Algorithm~\ref{alg:insert} which allows us to train our blind adversarial perturbation model. We define the gradient function for different types of feature as described in the following.



\begin{algorithm}[t!]
\begin{footnotesize}
    \caption{Packet insertion remapping function}
   \begin{algorithmic}
       \STATE $l \gets G(z)$
       \STATE $\bm{x} \gets $ input
       \STATE $n \gets$ number of added packets
       \STATE $p =$ position of top $n$ absolute   values of $l$
       \FOR{ $i$ in $p$}
       \STATE insert $+1$ if $l[i]>0$, otherwise $-1$ to $x$ at position $i$ and shift other features
       \ENDFOR
       \RETURN $\bm{x}$
    \end{algorithmic}
    \label{alg:insert}
\end{footnotesize}
\end{algorithm}

\paragraphb{Injecting adversarial  directions:}
While an adversary cannot change the directions of existing packets, she can inject adversarial directions by adding packets.
A connection's packet directions can be represented as  a series of -1 (downstream) and +1 (upstream) values. However, generating adversarial perturbations with binary values is  not straightforward.

We generate a   perturbation vector $G(z)$ with the same size as the target connection.
 Each element of this vector shows the effect of inserting a packet at that specific position (i.e., $l$ in Algorithm~\ref{alg:insert}). We select  positions with  largest absolute values for packet injection; the sign of the selected position determines the direction of the injected packet.
%
Finally, we modify the perturbation model's gradient as:
\begin{align}\label{eq:grad_where}
	\nabla_{G(z)} = \sum_{\bm{x} \in b_i}\nabla_{\bm{x}} {\mathcal{M}^I(\bm{x},G(z))}
\end{align}

\paragraphb{Injecting adversarial  timings/sizes:} Unlike packet directions, for the timing and size features, we need to learn both the positions and the values of the added packets simultaneously.  We design the perturbation generation model to output two vectors for the locations and the values of the added packets, where the value vector represents the selected feature (timing or sizes).
We use the gradient function defined in \eqref{eq:grad_where} for the position of the inserted packets. We use Algorithm~\ref{alg:grad_v} to compute the gradients for the values of the inserted packets.

\begin{algorithm}[t!]
\begin{footnotesize}
    \caption{Value Vector Gradient}
   \begin{algorithmic}
       \STATE $l,a \gets G(z)$
       \STATE $\nabla \mathcal{M}(\bm{x},G(z)) \gets $ gradient w.r.t. $\mathcal{M}(\bm{x},G(z))$
       \STATE $\nabla G(z) \gets \vv{\bm{0}}$
       \STATE $n \gets$ number of added packets
       \STATE $p =$ position of top $n$ values of $l$
       \FOR{ $i$ in $p$}
       \STATE $\nabla G(z)[i]=\nabla \mathcal{M}(x,G(z))[i]$
       \ENDFOR
       \RETURN $\nabla G(z)$
    \end{algorithmic}
    \label{alg:grad_v}
\end{footnotesize}
\end{algorithm}

\paragraphb{Injecting  multiple adversarial  features:}
To inject packets that simultaneously perturb several features, we  modify the perturbation generation model $G$  to output one vector for the position of the injected packets and one for each feature set to be perturbed.
We use Algorithm~\ref{alg:grad_v} to compute the gradient of each  vector. Moreover, we cannot use \eqref{eq:grad_where} to compute the gradient for the position vector, therefore, we take the average between the gradient of all different input feature vectors.

\section{Experimental Setup}\label{sec:setup}

Here we discuss the setting of our experiments, which we implemented in PyTorch~\cite{paszke2017automatic}.

\subsection{Metrics}
%
For a given blind adversarial perturbation generator $G(\cdot)$ and test dataset $\mathcal{D}_{test}$,
we define the \emph{attack success} metric as:
\begin{align}\label{eq:success}
  \mathcal{A} =
  \begin{cases}
     \frac{1}{|\mathcal{D}_{test}|} \sum_{(\bm{x},y) \in \mathcal{D}_{test}}  \mathbbm{1}[f(\bm{x}+G(z))\neq y] ~~~ &\code{DU}\\
   \frac{1}{|\mathcal{D}_{test}|} \sum_{(\bm{x},y) \in \mathcal{D}_{test}}  \mathbbm{1}[f(\bm{x}+G(z))=t] ~~~ &\code{DT}\\
 \end{cases}
\end{align}
where \code{DU} and \code{DT} represent destination-untargeted and targeted  attack scenarios, respectively (as defined in Section~\ref{sec:threat}).
For source-targeted (\code{ST}) cases, $\mathcal{D}_{test}$ contains instances only from the target source class.
Also, in our evaluations of the targeted attacks (\code{ST} and \code{DT}), we only  report the results for target classes with minimum and maximum attack accuracies. For example, ``Max \code{ST}-\code{DT}'' indicates the best results for the source and destination targeted attacks,  and we present the target classes using the  $TargetDest \leftarrow TargetSrc$ notation, which means class $TargetDest$ is  the targeted destination class and $TargetSrc$ is the targeted source class.
The maximum accuracy shows the worst case scenario for the target model and the minimum accuracy shows the lower bound on the adversary's success rate.
If there are multiple classes that lead to a max/min accuracy, we only mention one of them.

Note that while we can use $\mathcal{A}$ to evaluate attack success in various settings, for the  flow correlation experiments we use a more specific metric (as there are only two output classes for a flow correlation classifier). Specifically,  we use the  reduction in true positive and false positive rates of the target flow correlation algorithm to evaluate the success of our attack.

\subsection{Target Systems}\label{sec:target}
We demonstrate our attack on three  state-of-the-art DNN-based traffic analysis systems. 



\paragraphb{DeepCorr:}  DeepCorr~\cite{deepcorr} is the state-of-the-art flow correlation system, which uses   deep learning to learn flow correlation functions for specific network settings like that of Tor.
DeepCorr uses inter-packet delays (IPDs) and sizes of the packets as the features.
DeepCorr uses Convolutional neural networks to extract complex features from the raw timing and size information, and it outperforms the conventional statistical flow correlation techniques by significant margins.
Since DeepCorr uses both timings and sizes of packets as the features, we apply the time-based and size-based attacks on DeepCorr.

As mentioned earlier, non-blind adversarial perturbations~\cite{mockingbird, zhang2019statistical} are useless in the flow correlation setting, as the adversary does not know the features of the upcoming packets in a target connection. Hence, our blind perturbations are applicable in this  setting.

\paragraphb{Var-CNN:}
Var-CNN~\cite{var-cnn} is a deep learning-based website fingerprinting (WF) system that uses both manual and automated feature extraction techniques to be able to work with even  small amounts of training data.
Var-CNN uses ResNets~\cite{He2015DeepRL} with dilated casual convolutions, the state-of-the-art convolutional neural network, as its base structure.
Furthermore, Var-CNN shows that in contrast to previous WF attacks, combining packet timing information (IPDs) and direction information can improve the performance of the WF adversary.
In addition to packet IPDs and directions, Var-CNN uses cumulative statistical information for features of network flows.
Therefore, Var-CNN combines three different models, two ResNet models for timing and direction information, and one fully connected model for metadata statistical information as the final structure.
Var-CNN considers both \textit{closed-world} and \textit{open-world} scenarios. 

Similar to the setting of flow correlation, a WF adversary will not be able to use traditional (non-blind) adversarial perturbations~\cite{mockingbird, zhang2019statistical}, as she will not have knowledge on the patterns of upcoming packets in a targeted connection. Therefore, WF is a trivial application for blind perturbations.
 Since Var-CNN uses both IPD and packet direction features for fingerprinting, we use both timing-based and direction-based techniques  to generate our adversarial perturbations.

\paragraphb{Deep Fingerprinting (DF):}
Deep Fingerprinting (DF)~\cite{sirinam2018deep} is a deep learning based WF attack which uses CNNs to perform  WF attacks on Tor.
DF deploys automated feature extraction, and uses the direction information for training. In contrast to Var-CNN, DF does not require handcrafted features of packet sequences.
Similar to Var-CNN, DF considers both closed-world and open-world scenarios. Sirinam et al.~\cite{sirinam2018deep} show that DF outperforms prior WF systems in defeating  WF defenses  of WTF-PAD~\cite{juarez2016toward} and W-T~\cite{wang2017walkie}.

\paragraphb{Codes.} As we perform our attack in PyTorch, we use the original codes of DeepCorr, DF, and Var-CNN models and convert them from TensorFlow to PyTorch.
We then train these models using the  datasets of those papers.

\subsection{Adversary Setup and Models}




While our technique can be applied to any traffic analysis setting, we present our setup for the popular Tor application.

\paragraphb{Adversary's Interception Points}
Our adversary has the same placement as traditional Tor traffic analysis works~\cite{sirinam2018deep,wang2017walkie,rimmer2017automated, wang2013improved, wang2016realistically}.
For the WF scenario, we assume the adversary is manipulating the traffic between a Tor client and the first Tor hop, i.e., a Tor bridge~\cite{tor-bridge} or a Tor guard relay.
Therefore, our blind adversarial perturbation can be implemented as a \emph{Tor pluggable transport}~\cite{plugtor},
in which case the blind perturbations are applied by both the Tor client software and the Tor bridge.
In the flow correlation setting, similar to the literature, traffic manipulations are performed by Tor entry and exit relays (since flow correlation attackers intercept  both egress and ingress Tor connections).
In our evaluations, we show that even applying our blind adversarial perturbations on only ingress flows is enough to defeat flow correlation attacks, i.e., the same adversary placement as the WF setting.


\paragraphb{Adversarial Perturbation Models}
As mentioned in Section~\ref{sec:method}, we design a deep learning model to generate blind adversarial noises.
For each type of perturbation, the adversarial model is a fully connected model with one hidden layer of size 500 and a ReLu activation function.
The parameters of the adversarial model are presented in Appendix~\ref{app:params}.
The input and output sizes of the adversarial model are equal to the length of features in the target flow.
In the forward function, the adversarial model takes in a given input, manipulates it based on the attack method, and output a crafted version of the input.
In each iteration of training, we update the  parameters of the adversarial model based on the loss functions introduced in Section~\ref{sec:method}. We use Adam optimizer to learn the blind noise with a learning rate of $0.001$.

\paragraphb{Discriminator Model}
As  mentioned in Section~\ref{sec:method}, we use a  GAN model to enforce the time constraints of our modified network flows.
To do so, we design a fully-connected discriminator model containing two hidden FC layers of size 1000.
The parameters of the discriminator model are presented in Appendix~\ref{app:params}.
The input and output sizes of this model are equal to the sizes of the blind adversarial noise. In the training process, we use Adam optimizer with a learning rate of $0.0001$ to learn the discriminator model.

\subsection{Datasets}

\paragraphb{Tor Flow Correlation Dataset}
For flow correlation experiments, we use the publicly available dataset of DeepCorr~\cite{deepcorr}, which contains $7000$ flows for training and $500$ flows for testing. These flows are captured Tor flows of top Alexa's websites and contain timings and sizes of each of them.
These flows are then used to create a large set of flow pairs including associated flow pairs (flows belonging to the same Tor connection) and non-associated flow pairs (flows belonging to arbitrary Tor connections). Each associated flow pair is labeled with $1$, and each non-associated flow pair with $0$.


\paragraphb{Tor Website Fingerprinting Datasets}
Var-CNN uses a dataset of 900 monitored sites each with 2,500 traces. These sites were compiled from the Alexa list of most popular websites. Var-CNN is fed in with different sets of features representing a given trace;
the direction-based ResNet model takes a set of 1's and -1's as the direction of each packet such that 1 shows an outgoing packet and -1 represents an incoming packet. The time-based ResNet uses the IPDs of the traces as features. The metadata model takes in seven float numbers as the statistical information of the traces. To be consistent with previous WF attacks~\cite{sirinam2018deep, rimmer2017automated, wang2013improved, wang2016realistically}, we use the first 5000 values of a given trace for both direction and time features.

DF uses a different dataset than Var-CNN. For the closed-world setting, they collected the traces of 95 top Alexa websites with 1000 visits for each. DF uses the same representation as Var-CNN for direction information of the packets. Since CNNs only take in a fixed length input, DF considers the first 5000 values of each flow.




\section{Experiment Results}


We  evaluate  blind adversarial perturbations against the target systems of Section~\ref{sec:target} using  each of the three key  traffic features and their combinations. We also compare our attack with traditional attacks.


\subsection{Adversarially Perturbing  Directions}

\begin{table*}[t]
   \centering
   \caption{Direction perturbation attack on DeepFingerprinting~\cite{sirinam2018deep} WF scheme ($92\%$  WF accuracy)}
   \resizebox{\textwidth}{!}{
\begin{tabular}{@{}l|l|llllllll@{}}
    \toprule
$\alpha$ &Bandwith Overhead ($\%$) &  $\mathcal{A}:$& \code{SU-DU}  (\%) & Max \code{ST-DU} (\#, $\%$)  & Min \code{ST-DU}  (\#, $\%$) & Max \code{SU-DT} (\#, $\%$) & Min \code{SU-DT} (\#, $\%$) & Max \code{ST-DT} (\#$\leftarrow $\#, $\%$) & Min \code{ST-DT} (\#$\leftarrow $\#, $\%$)\\
\midrule
20 & 0.04 && $24.2 $  & $-, 100.0$ & $-, 0.0$     & $77, 31.9$  & $4, 0.1$  &  $-, 100.0$  & $-, 0.0 $  \\
100 & 2.04 &&$49.6 $ & $-, 100.0$ & $47, 0.0$ & $34, 77.6$ & $89, 13.2$ &  $-, 100.0$     & $-, 0.0 $ \\
500 & 11.11 &&$91.8 $ & $-, 100.0$ & $49, 4.0$ & $92, 97.1$& $82, 47.8$ &  $-, 100.0$   & $23\leftarrow 69, 0.1 $  \\
1000 & 25.0 && $95.7 $ & $-, 100.0$ & $21, 29.0$ & $-, 100.0$& $10, 67.0$ &  $-, 100.0$   & $72\leftarrow 47, 4.4 $  \\
2000 & 66.66 && $97.7 $ & $-, 100.0$ & $48, 94.7$ & $-, 100.0$& $37, 89.4$ &  $-, 100.0$   & $78\leftarrow 60, 35.4 $  \\
\bottomrule
\end{tabular}}
\label{tab:wf_res_deep}
\end{table*}

As explained in Section~\ref{sec:method}, an adversary cannot change the  directions of existing packets, but he can  insert packets with adversarial  directions.
We evaluated our attack for different adversary settings and strengths against Var-CNN~\cite{var-cnn} and DF~\cite{sirinam2018deep} (which use direction features).
We used 10 epochs and  Adam optimizer to train the blind adversarial perturbations model with a learning rate of $0.001$.
Tables~\ref{tab:wf_res_deep} and~\ref{tab:wf_res_var} show the success of our attack (using $\mathcal{A}$ in \eqref{eq:success})  on DF and Var-CNN,  respectively,  when they only use  packet directions as their features.
As can be seen, \textbf{both DF and Var-CNN are highly vulnerable to adversarial perturbation attacks} when the adversary only injects a small  number of  packets.
Specifically, we were able to generate targeted perturbations that  misclassify \emph{every} input into a target class with only 25$\%$ bandwidth overhead.


\subsection{Adversarially Perturbing  Timings}

We consider two  scenarios for  generating adversarial timing perturbations: with and without an invisibility constraint. In both scenarios, we limited the adversaries' power such that the added noise to the timings of the packets has a  maximum mean and standard deviation as explained in Section~\ref{sec:method}. For the invisibility constraint, we force the added noise to have the same distribution as natural network jitter, which follows a  Laplace distribution~\cite{deepcorr}. The detailed parameters of our model are presented in Appendix~\ref{app:params}.

 \begin{table*}[t!]
   \centering
   \caption{Direction perturbation attack on Var-CNN~\cite{var-cnn} WF scheme ($93\%$ WF accuracy)}
      \resizebox{\textwidth}{!}{
\begin{tabular}{@{}l|l|llllllll@{}}
    \toprule
$\alpha$ &Bandwith Overhead ($\%$) &  $\mathcal{A}:$ & \code{SU-DU}  (\%) & Max \code{ST-DU} (\#, $\%$)  & Min \code{ST-DU}  (\#, $\%$) & Max \code{SU-DT} (\#, $\%$) & Min \code{SU-DT} (\#, $\%$) & Max \code{ST-DT} (\#$\leftarrow $\#, $\%$) & Min \code{ST-DT} (\#$\leftarrow $\#, $\%$)\\
\midrule
20 & 0.04 && $76.1 $  & $-, 100.0$ & $-, 0.0$     & $2, 68.3$  & $8, 53.2$  &  $-, 100.0$  & $-, 0.0 $  \\
100 & 2.04 && $80.3 $ & $-, 100.0$ & $-, 100.0$ & $4, 76.5$ & $2, 66.8$ &  $-, 100.0$     & $-, 0.0 $ \\
500 & 11.11   &&$96.8 $ & $-, 100.0$ & $-, 100.0$ & $3, 98.9$& $9, 81.7$ &  $-, 100.0$   & $-, 10.0 $  \\
1000 & 25.0 && $98.2 $ & $-, 100.0$ & $-, 100.0$ & $-, 100.0$& $0, 96.6$ &  $-, 100.0$   & $-, 20.0 $  \\
2000 & 66.66 && $99.0 $ & $-, 100.0$ & $-, 100.0$ & $-, 100.0$& $8, 97.6$ &  $-, 100.0$   & $-, 30.0 $  \\
\bottomrule
\end{tabular}}
\label{tab:wf_res_var}
\end{table*}

\begin{figure*}[t]
\centering
   \begin{subfigure}{0.32\textwidth}
       \centering
		\includegraphics[width=1.\textwidth]{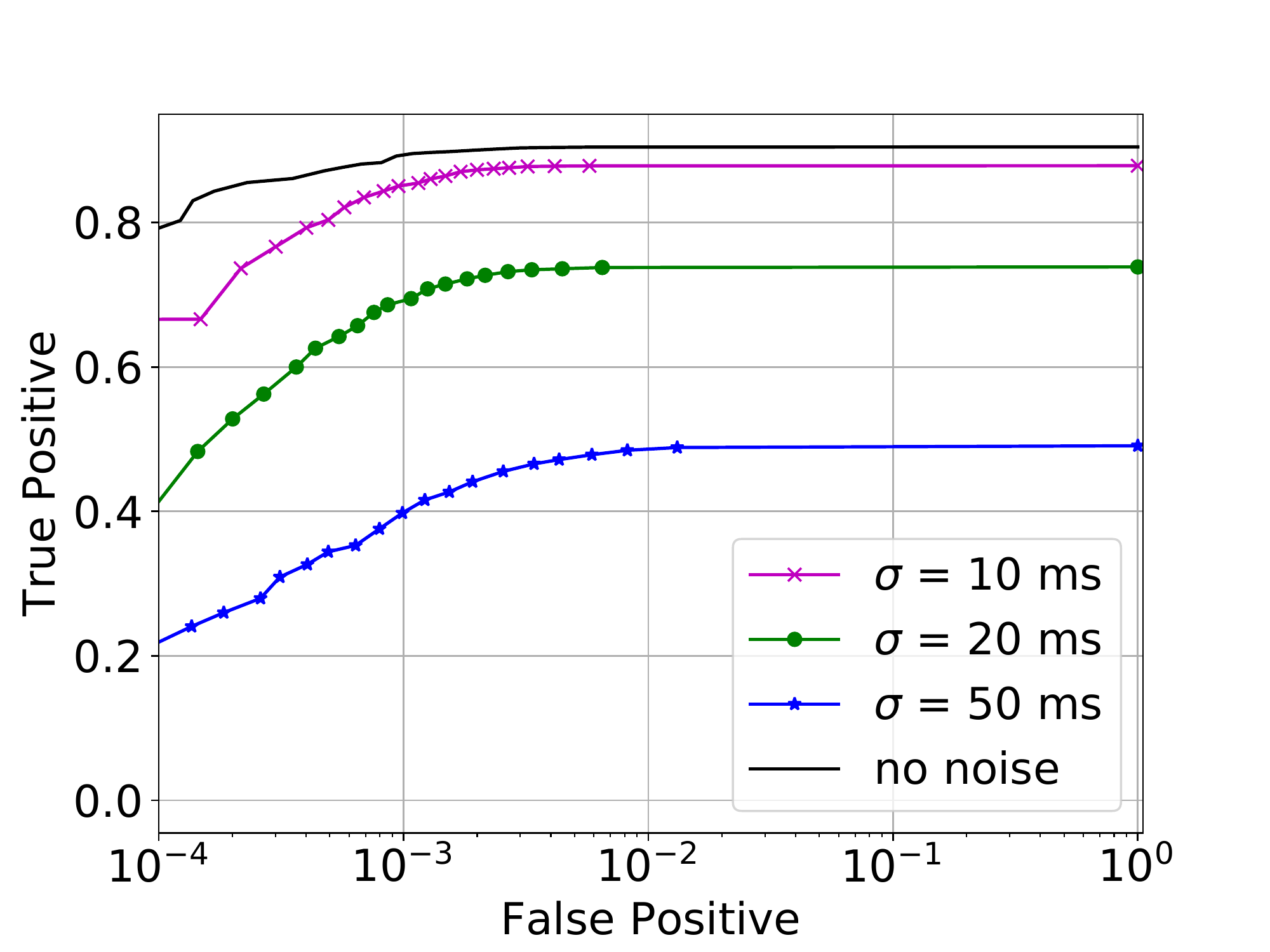}
		\caption{$\mu =0 ms $}
    \label{fig:deepcorr_eps}
	\end{subfigure}
	\begin{subfigure}{0.32\textwidth}
       \centering
		\includegraphics[width=1.\textwidth]{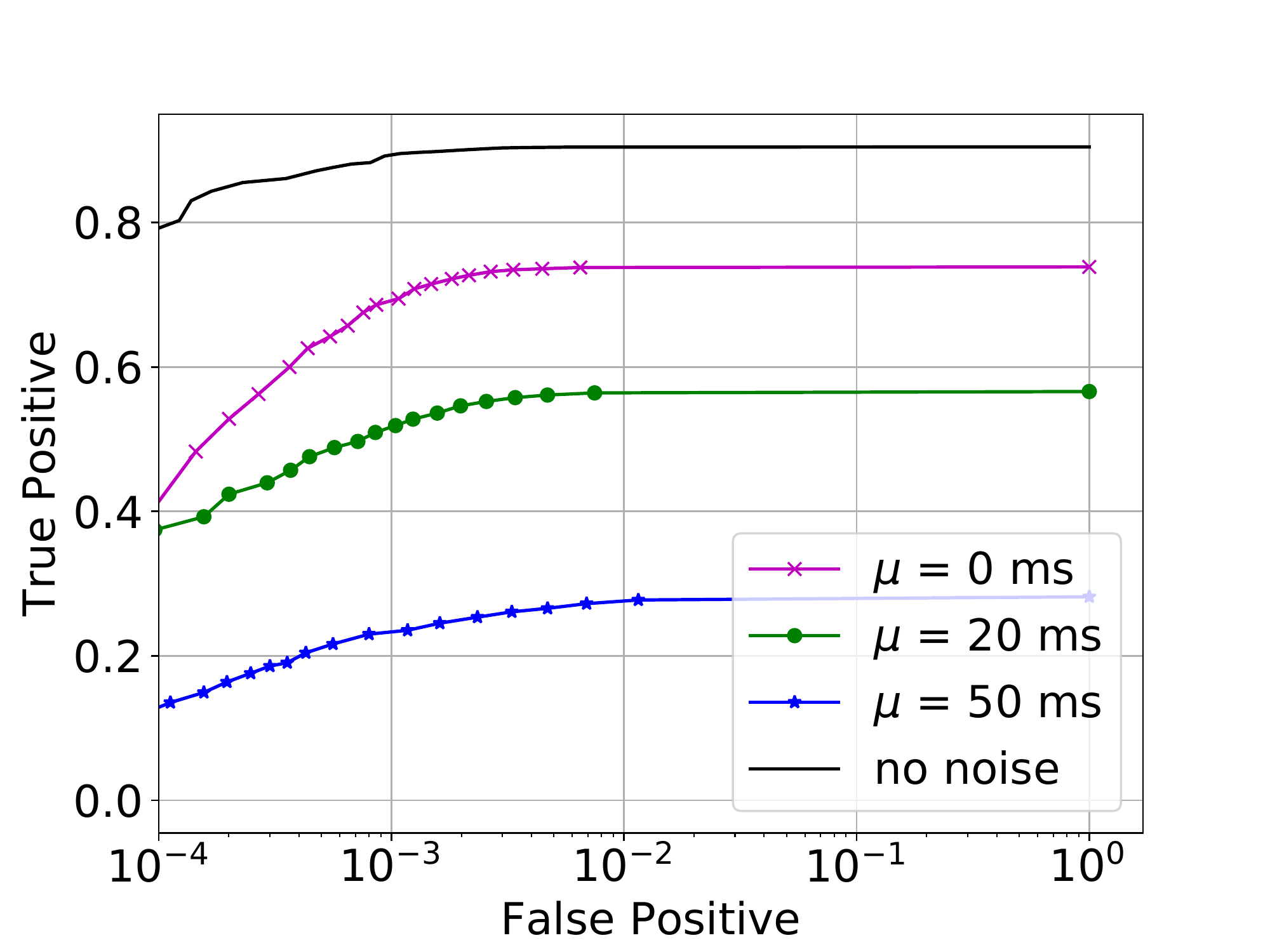}
		\caption{$\sigma =20 ms$}
    \label{fig:deepcorr_mid}
	\end{subfigure}
	\begin{subfigure}{0.32\textwidth}
       \centering
		\includegraphics[width=1.\textwidth]{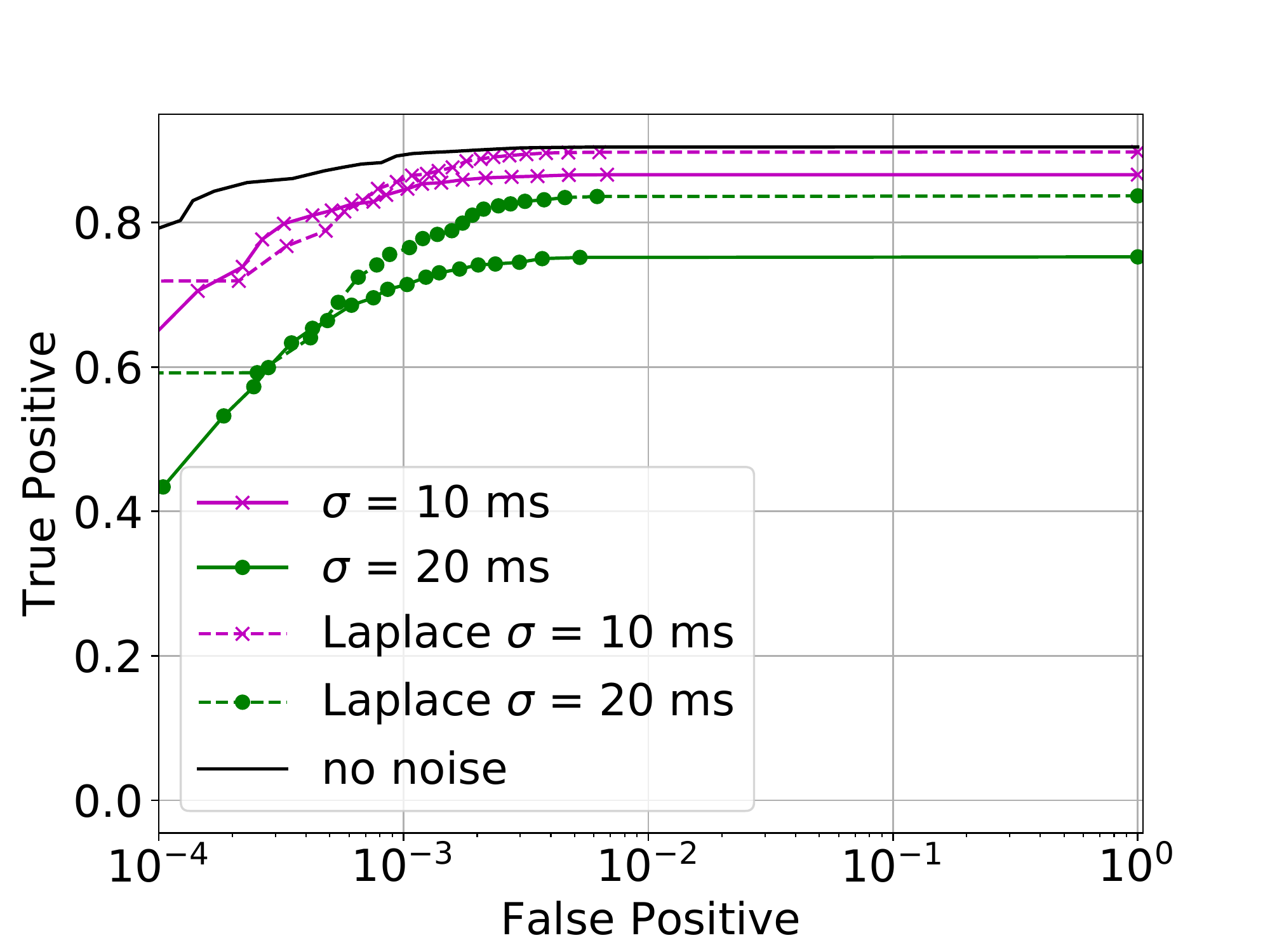}
		\caption{$\mu =0 ms $ with invisibility constraint}
    \label{fig:deepcorr_laplace}
	\end{subfigure}
	\caption{Timing perturbations on  DeepCorr for different attack strengths, with/without an invisibility constraint.}
	\label{fig:deepcorr_timing}
\end{figure*}

Figure~\ref{fig:deepcorr_timing} shows the performance of our attack against DeepCorr when the adversary only manipulates the timings of  packets. As expected, Figures~\ref{fig:deepcorr_eps} and \ref{fig:deepcorr_mid} show  that increasing the strength (mean or standard deviation) of our blind noise results in better performance of the attack, but  \textbf{even a perturbation with average  0 and  a tiny standard deviation of $50ms$ significantly reduces  the true positive of  DeepCorr from $96\%$ to $40\%$}.

\begin{figure*}[!t]
\centering
\minipage{0.31\linewidth}
\centering
	\includegraphics[width = \linewidth]{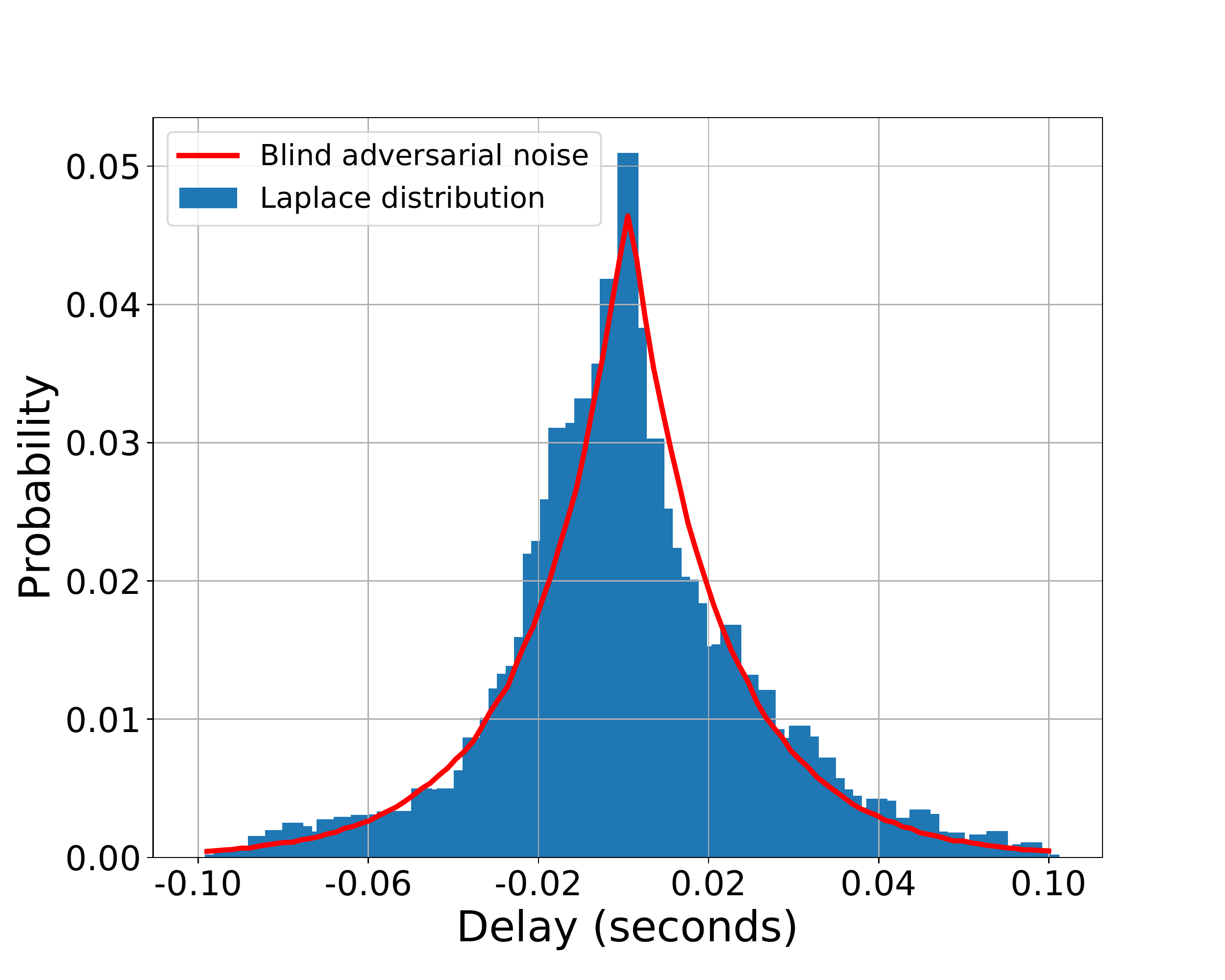}
	\caption{Blind timing perturbations generated to follow a Laplace distribution with $\mu=0, \sigma=30ms$.}
	\label{fig:laplace}
\endminipage\hfill
\minipage{0.33\linewidth}
	\centering
	\includegraphics[width = \linewidth]{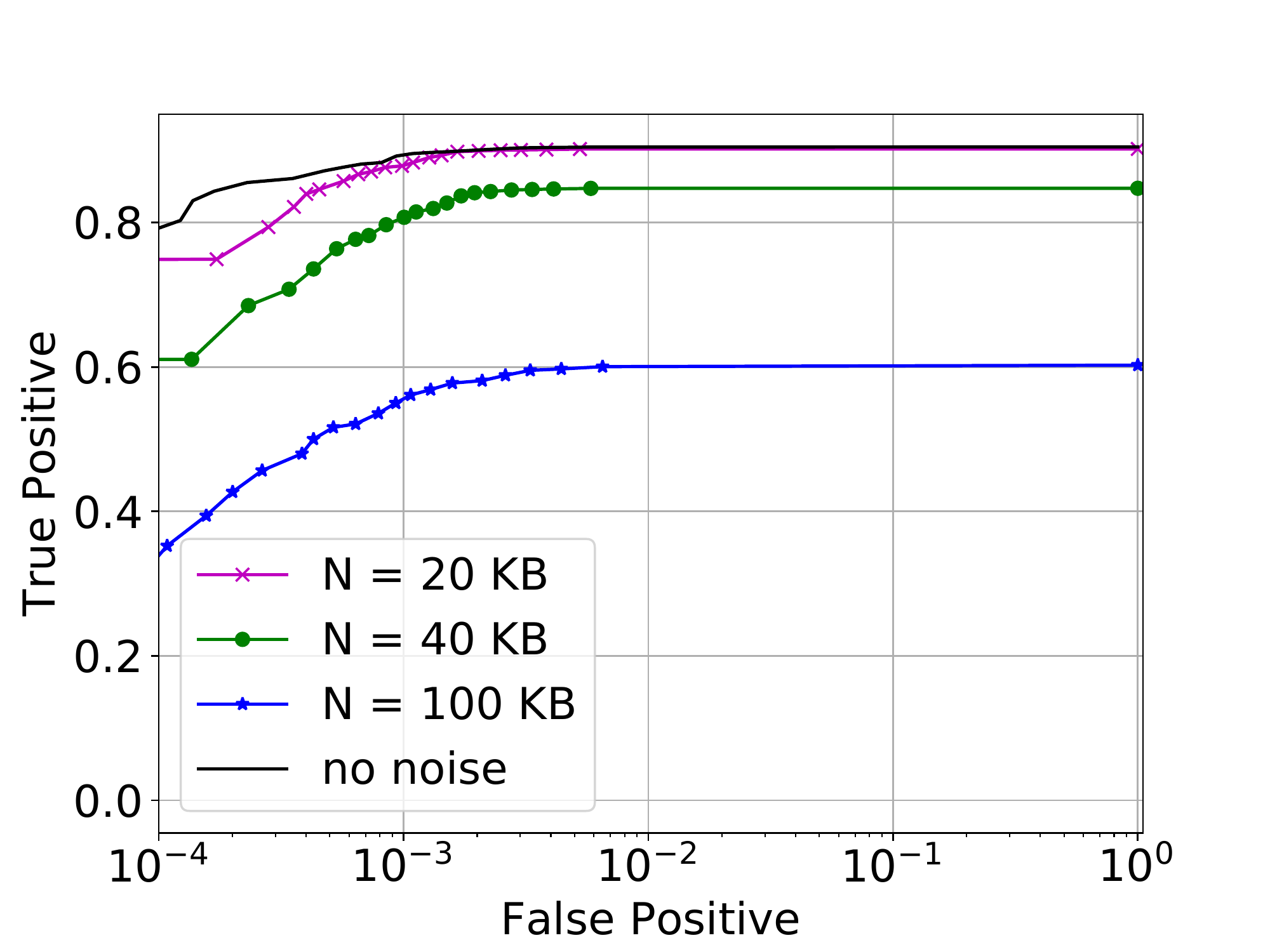}
	\caption{Size perturbations on  DeepCorr for different attack strengths}
	\label{fig:deepcorr_size}
\endminipage\hfill
\minipage{.31\linewidth}
\centering
	\includegraphics[width = \linewidth]{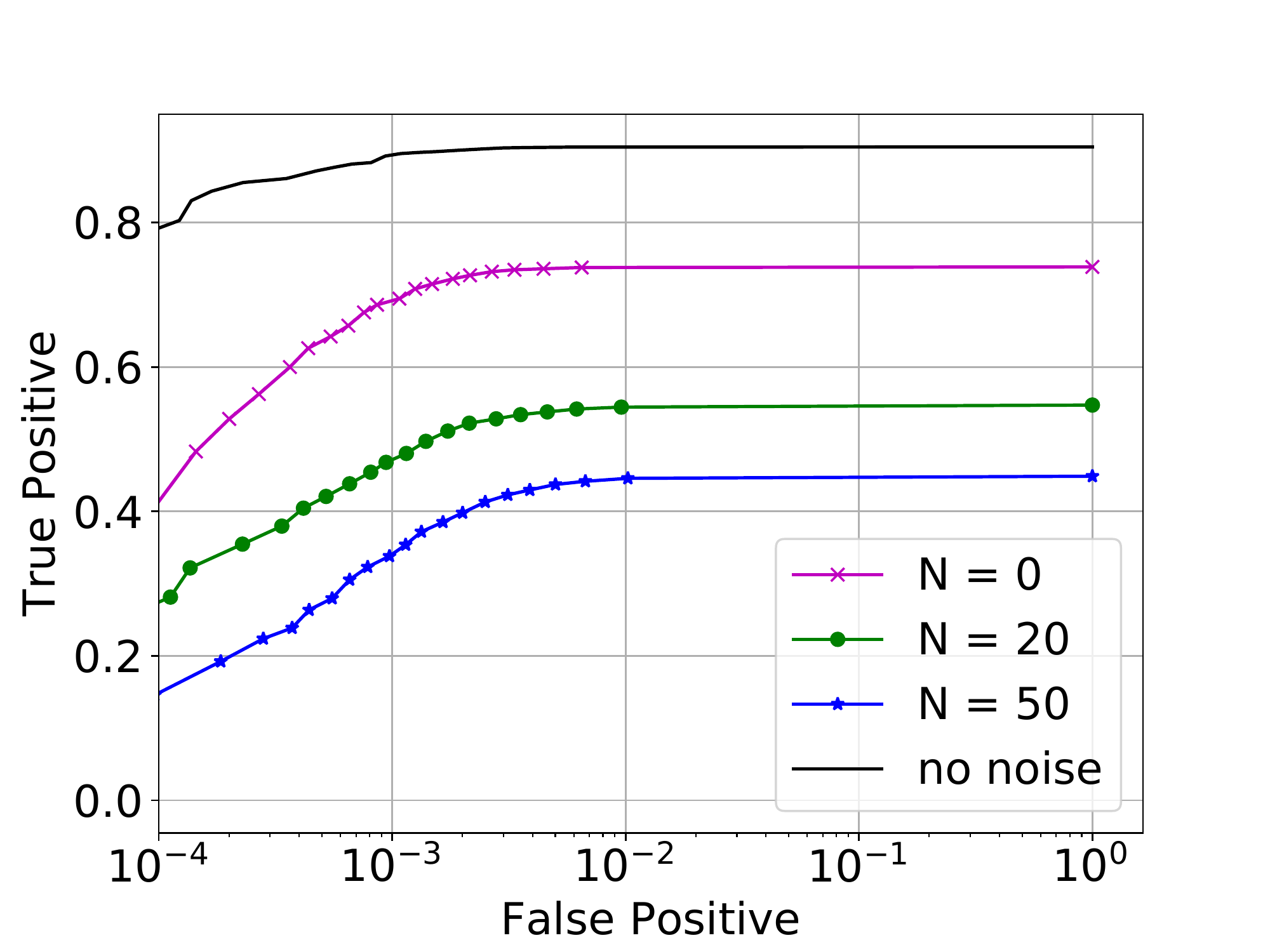}
	\caption{Hybrid size/timing perturbations on  DeepCorr for different attack strengths}
	\label{fig:deepcorr_all}
\endminipage\hfill
\end{figure*}

Also, \textbf{we can create effective adversarial perturbations with high invisibility}: Figure~\ref{fig:laplace} shows the histogram of the generated timing perturbations, with parameters $\mu = 0, \sigma = 30ms$, learned under an invisibility constraint  forcing  it to follow a Laplace distribution.
For this invisible noise,   Figure~\ref{fig:deepcorr_laplace} compares the performance of timing perturbations on DeepCorr with different attack strengths; it  also  shows the impact of  arbitrary Laplace distributed perturbations on DeepCorr.

We also  apply our timing perturbations  on Var-CNN.
Table~\ref{tab:wf_timing_var} shows our attack success ($\mathcal{A}$)  with and without an invisibility constraint.
We realize that timing perturbations  have much larger impacts on Var-CNN than direction perturbations.
Moreover, as expected, in the untargeted scenario (\code{SU}-\code{DU}) and for different bandwidth overheads, our attack has better performance without the invisibility constraint.
However, \emph{even with an  invisibility constraint, our attack reduces the accuracy of Var-CNN drastically}, i.e., \textbf{a blind timing perturbation  with an average 0 and a tiny standard deviation of $20ms$ reduces the accuracy of Var-CNN by $89.6\%$}.

 \begin{table*}[t]
   \centering
   \caption{Timing perturbation attack on Var-CNN~\cite{var-cnn} WF scheme ($93\%$ WF accuracy)}
   \resizebox{\textwidth}{!}{
\begin{tabular}{@{}lllllllllll@{}}
    \toprule
    &&  \multicolumn{4}{c}{Limited Noise} &   \multicolumn{4}{c}{Stealthy Noise} \\
      \cmidrule{3-6} \cmidrule{8-11}  $\mu,\sigma$ & $\mathcal{A}:$ & \code{SU-DU}  ($\%$) & Max \code{ST-DU} (\#, $\%$)   & Max \code{SU-DT} (\#, $\%$) &  Max \code{ST-DT} (\#$\leftarrow $\#, $\%$) && \code{SU-DU}  ($\%$) & Max \code{ST-DU} (\#, $\%$)   & Max \code{SU-DT} (\#, $\%$) &  Max \code{ST-DT} (\#$\leftarrow $\#, $\%$) \\
\midrule
0,5 && $37.7$ & $100.0 $  & $17, 40.3$ & $-, 100.0$     && $22.7$  & $100.0$  &  $17, 40.3$  & $-, 100.0 $  \\
0,10 && $68.1$ & $100.0 $  & $53, 83.4$ & $-, 100.0$     && $38.7$  & $100.0$  &  $53, 83.4$  & $-, 100.0 $  \\
0,20 && $96.6$ & $100.0 $  & $80, 95.8$ & $-, 100.0$     && $91.5$  & $100.0$  &  $80, 95.8$  & $-, 100.0 $  \\
0,30 && $94.3$ & $100.0 $  & $80, 99.7$ & $-, 100.0$     && $90.4$  & $100.0$  &  $80, 99.7$  & $-, 100.0 $  \\
0,50 && $98.8$ & $100.0 $  & $80, 100.0$ & $-, 100.0$     && $97.9$  & $100.0$  &  $80, 100.0$  & $-, 100.0 $  \\
\bottomrule
\end{tabular}}
\label{tab:wf_timing_var}
\end{table*}

\subsection{Adversarially Perturbing Sizes}

 We evaluate  our size perturbation attack  on DeepCorr, which  is the only system (among the three we studied) that uses packet sizes for traffic analysis.
As DeepCorr is mainly studied in the context of Tor, our perturbation algorithm enforces the size distribution of  Tor on the generated size perturbations.
Figure~\ref{fig:deepcorr_size} shows the results when the adversary only manipulates  packet sizes.
As can be seen, size perturbations are less impactful on  DeepCorr than timing perturbations, suggesting that DeepCorr is more sensitive to the timings of  packets.

\subsection{Perturbing Multiple Features}

In this section, we evaluate the performance of our adversarial perturbations when we perturb multiple features simultaneously.
Var-CNN uses both packet timing and directions to fingerprint websites.  Table~\ref{tab:wf_res_var_all} shows the impact of adversarially perturbing both of these features on Var-CNN; we see that  \textbf{combining perturbation attacks increases the impact of the attack}, e.g., in the untargeted scenario (\code{SU-DU}), the combination of both attacks with parameters $\alpha = 100, \mu = 0, \sigma = 10ms$ results in an attack success of  $\mathcal{A}=83.9\%$  while the time-based and direction-based perturbations alone result in  $\mathcal{A}=68.1\%$  and $\mathcal{A}=80.3\%$, respectively.
Similarly, in Figure~\ref{fig:deepcorr_all}, we see that by combining time and size perturbations, the accuracy of DeepCorr drops from $88\%$ to $49\%$ (with $FP = 10^{-3}$) by injecting only 20 packets, while using only time perturbations the accuracy drops to $68\%$.

 \begin{table*}[t]
   \centering
   \caption{Hybrid time/direction perturbations on Var-CNN~\cite{var-cnn}.} 
   \resizebox{\textwidth}{!}{
\begin{tabular}{@{}l|l|llllllll@{}}
    \toprule
$\alpha, \mu, \sigma$,  &BW Overhead ($\%$) & $\mathcal{A}$: & \code{SU-DU}  (\%) & Max \code{ST-DU} (\#, $\%$)  & Min \code{ST-DU}  (\#, $\%$) & Max \code{SU-DT} (\#, $\%$) & Min \code{SU-DT} (\#, $\%$) & Max \code{ST-DT} (\#$\leftarrow $\#, $\%$) & Min \code{ST-DT} (\#$\leftarrow $\#, $\%$)\\
\midrule
20, 0, 5 & 0.04 & &$79.0 $  & $-, 100.0$ & $4, 30.0$ & $2, 69.4$  & $6, 40.3$  &  $-, 100.0$  & $-, 0 $  \\
100, 0, 10 & 2.04 &&$83.9 $ & $-, 100.0$ & $-, 100.0$ & $2, 92.8$ & $3, 72.3$ &  $-, 100.0$     & $-, 10.0 $ \\
500, 0, 20 & 11.11 &&$97.0 $ & $-, 100.0$ & $-, 100.0$ & $3, 99.9$& $4, 92.6$ &  $-, 100.0$   & $-, 20.0 $  \\
1000, 0, 30 & 25.0 & &$98.6 $ & $-, 100.0$ & $-, 100.0$ & $-, 100.0$& $0, 96.7$ &  $-, 100.0$   & $-, 30.0 $  \\
2000, 0, 50 & 66.66 & &$99.0 $ & $-, 100.0$ & $-, 100.0$ & $-, 100.0$& $9, 97.7$ &  $-, 100.0$   & $-, 30.0 $  \\
\bottomrule
\end{tabular}}
\label{tab:wf_res_var_all}
\end{table*}

\subsection{Comparison With Traditional Attacks }\label{sec:comp-defense}

There exist traditional attacks on DNN-based traffic analysis systems that use techniques other than adversarial perturbations. In this section, we compare our adversarial perturbation attacks with such traditional approaches.

\paragraphbe{Packet insertion techniques: } Several WF countermeasures work by adding new packets.
We show that \textbf{our adversarial perturbations are significantly more effective} with similar overheads.
 WTF-PAD~\cite{juarez2016toward} is a state-of-the-art technique  which adaptively adds dummy packets to Tor traffic to evade website fingerprinting systems.
Using WTF-PAD on the DF dataset  reduces the WF  accuracy to $3\%$ at the cost of a  $64\%$ bandwidth overhead.
Similarly, the state-of-the-art Walkie-Talkie~\cite{wang2017walkie} reduces DF's accuracy to  $5\%$ with a $31\%$ bandwidth overhead and a $36\%$ latency overhead~\cite{sirinam2018deep}.
On the other hand,  our injection-based targeted blind adversarial attack  reduces the detection accuracy to $1\%$ (close to random guess) \emph{with only a $25\%$ bandwidth overhead and no added latency} (using the exact same datasets).
To compare existing WF countermeasures with our results while using Var-CNN model, we refer to their paper~\cite{var-cnn} where WTF-PAD can decrease the accuracy of Var-CNN by $0.4\%$ (from $89.2\%$ to $88.8\%$) with $27\%$ bandwidth overhead.
However, according to Table~\ref{tab:wf_res_var}, with a similar bandwidth overhead (1000 inserted packets and $25\%$ overhead), \textbf{our attack reduces the accuracy by $91.6\%$ which significantly outperforms  WTF-PAD}. Our results suggest that,  \textbf{our blind adversarial perturbation technique  drastically outperforms traditional defenses against deep learning based website fingerprinting systems}.

\paragraphbe{Time perturbation techniques: }  Figure~\ref{fig:deepcorr_laplace} compares our technique with a naive countermeasure of adding random Laplacian noise to the timings of the packets.
We see that by adding a Laplace noise with zero mean and 20ms standard deviation, the accuracy of DeepCorr drops from $0.88$ TP (for $10^{-3}$ FP) to $0.78$ TP, but using our adversarial perturbation technique with the same mean and standard deviation, the accuracy drops to $0.68$ and $0.71$ without and with invisibility, respectively.

\paragraphbe{Non-blind adversarial perturbations: }
Two recent works~\cite{mockingbird, zhang2019statistical} use ``non-blind'' adversarial perturbations to defeat traffic analysis classifiers. As discussed earlier, we consider these techniques unusable in regular traffic analysis applications, as they can not be applied on live connections. 
Nevertheless, we show that our technique even outperforms these non-blind techniques;  
for instance, when DF is the target system, 
Mockingbird~\cite{mockingbird} reduces the accuracy of DF by $59.8\%$ with a $56.5\%$ bandwidth overhead (in full-duplex mode), while our direction-based blind perturbation technique reduces the accuracy of DF by much higher $91.8\%$ and with a much lower bandwidth overhead of $11.11\%$.



\section{Countermeasures}

In this section, we evaluate defenses against our blind adversarial perturbations.
We start by showing why our perturbations are hard to counter.
We will then borrow three countermeasure techniques from the image classification domain, and show that
\textbf{they perform poorly} against blind adversarial perturbations.
Finally, we will design a tailored, more efficient defense on blind adversarial perturbations.

\begin{figure}[!t]
\minipage{0.45\linewidth}
\includegraphics[width = \linewidth, trim = {1.5cm 1.5cm 1.5cm 1.5cm}]{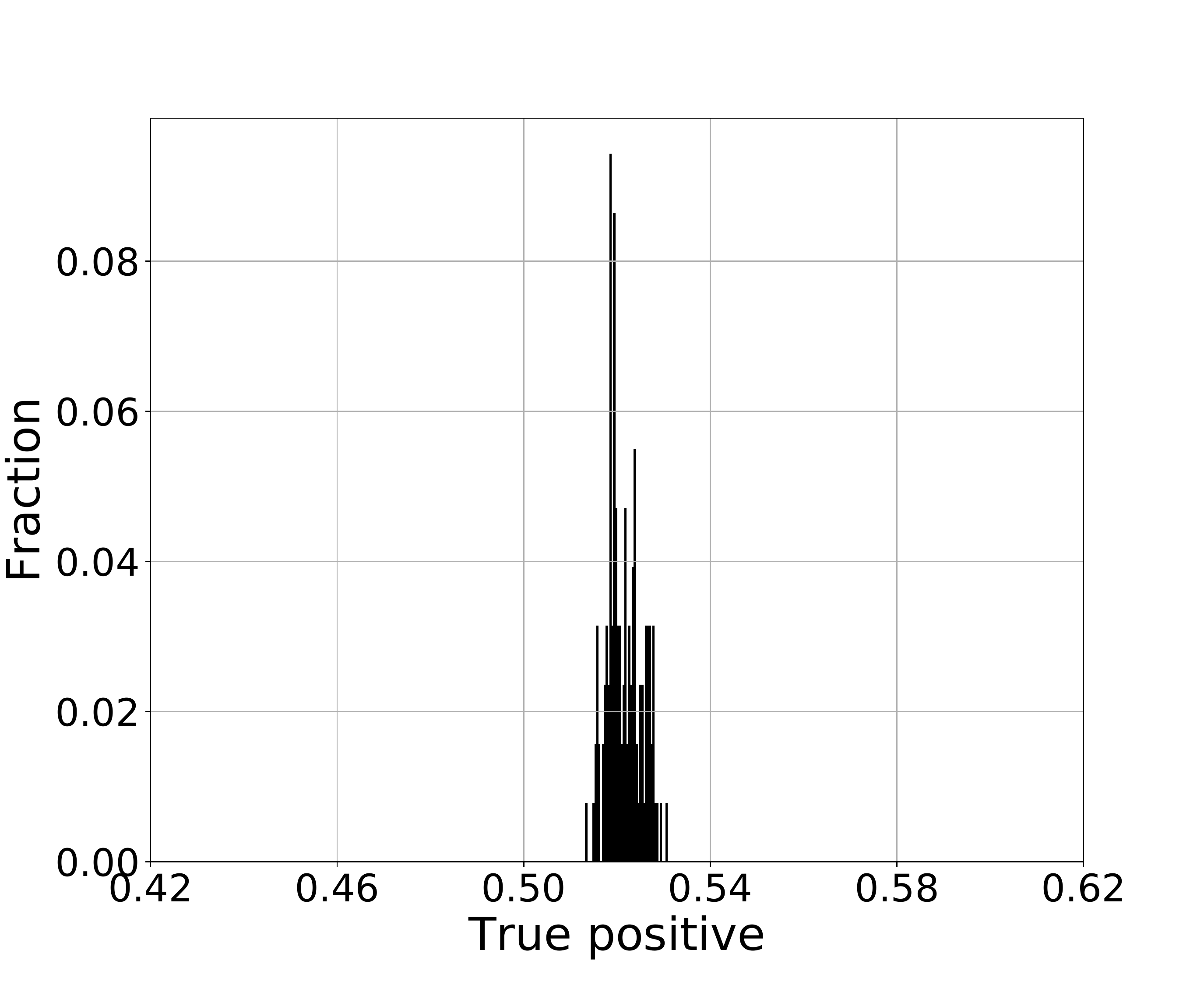}
     \caption{The accuracy of DeepCorr with different blind adversarial noises}
     \label{fig:acc-hist}
\endminipage\hfill
\minipage{0.45\linewidth}
\includegraphics[width = \linewidth, trim = {1.5cm 1.5cm 1.5cm 1.5cm}]{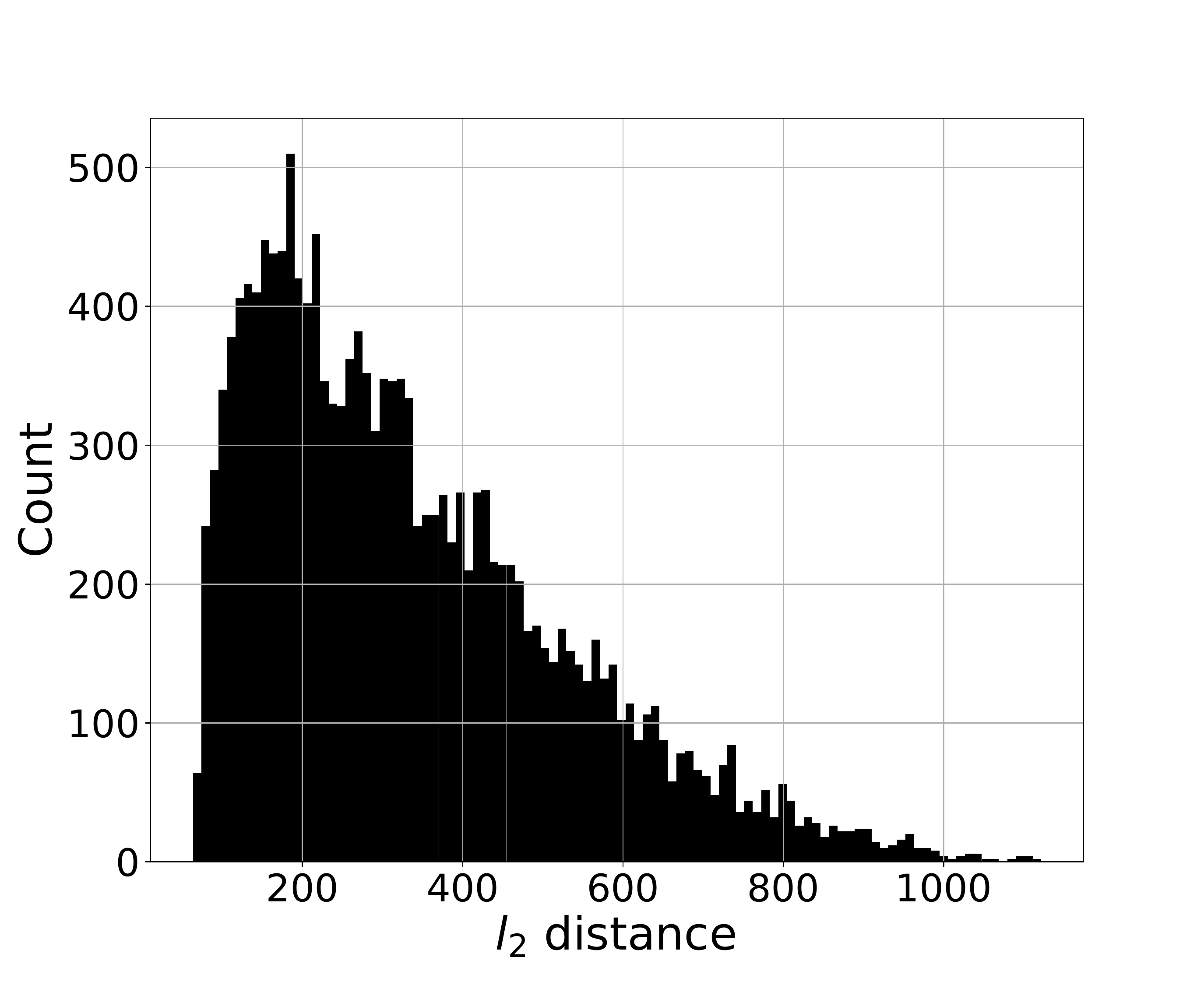}
     \caption{The $l_2$ distance between DeepCorr's different adversarial noises}
     \label{fig:l2-hist}
\endminipage\hfill
\end{figure}

\paragraphb{Uniqueness of Our Adversarial Perturbations:}
A key property that impacts countering adversarial perturbations is the \emph{uniqueness} of adversarial perturbations:
if there is only one (few) possible adversarial perturbations,  the defender can identify them and train her model to be robust against the known perturbations.
As explained before, our adversarial perturbations are not unique: our algorithm derives  a perturbation generator ($G(z)$) that for random $z$s can create different perturbation vectors.
To demonstrate the non-uniqueness of our perturbations, we  created 5,000 adversarial perturbations for the applications studied in this paper (we stopped at 5,000 only due to limited GPU memory). Figure~\ref{fig:l2-hist} shows the histogram of the $l2$ distance between the different adversarial perturbations that we generated for DeepCorr. We can say that  \textbf{the generated perturbations are not unique, and, the adversary cannot easily detect them}. These different  perturbations however cause similar adversarial impacts  on their target model, as shown in Figure~\ref{fig:acc-hist}.

\paragraphb{Adapting Existing Defenses:}
Many defenses have been designed for adversarial examples in   image classification applications, particularly,   \emph{adversarial training}~\cite{kurakin2016adversarial,tramer2017ensemble,madry2017towards}, \emph{gradient masking}~\cite{papernot2016distillation,ross2018improving},  and \emph{region-based classification}~\cite{cao2017mitigating}.
In Appendix~\ref{sec:prior-defense}, we discuss how we adapt these defenses to our networking application.

\begin{table}[t!]
   \centering
\caption{Evaluating various defenses against blind adversarial perturbations (website fingerprinting application)}   
   \resizebox{0.5\textwidth}{!}{
   \begin{tabular}{@{}lllllll@{}}
\toprule
Adversary Strength  & Original & No Def & Madry et al.~\cite{madry2017towards} & IGR~\cite{ross2018improving} & RC~\cite{cao2017mitigating} & Our Defense \\
\midrule
$\alpha = 20$ &  $92\%$ & $60\%$ & $84\%$ & $62\%$ & $54\%$ & $84\%$ \\
$\alpha = 100$ &  $92\%$ & $28\%$ & $48\%$ & $23\%$ & $23\%$ & $60\%$\\
$\alpha = 500$ & $92\%$ & $8\%$ & $19\%$ & $2\%$ & $7\%$ & $24\%$\\
\bottomrule
\end{tabular}}
\label{tab:defense_wf}
\vspace{0.2cm}
\end{table}

\begin{table}[t!]
   \centering
\caption{Evaluating various defenses against blind adversarial perturbations (flow correlation application).  FP=$10^{-4}$.}   
   \resizebox{0.5\textwidth}{!}{
   \begin{tabular}{@{}lllllll@{}}
\toprule
Adversary Strength & Original & No Def & Madry et al.~\cite{madry2017towards} & IGR~\cite{ross2018improving} & RC~\cite{cao2017mitigating} & Our Defense \\
\midrule
$\mu = 0, \sigma = 10$ & $79\%$ &   $63\%$ &  $70\%$ &  $62\%$ &  $63\%$ &  $74\%$\\
$\mu = 0, \sigma = 50$ & $79\%$ & $21\%$ &   $25\%$ &  $23\%$ &  $22\%$ &  $32\%$\\
$\mu = 0, \sigma = 100$ & $79\%$ & $13\%$  & $18\%$ &  $13\%$&  $14\%$ &  $23\%$\\
\bottomrule
\end{tabular}}
\label{tab:defense_deep}
\end{table}

\begin{table*}[t!]
   \hfill
   \begin{minipage}{0.3 \linewidth}
   \centering
   \caption{Transferability of direction-based  perturbations (surrogate model: DF~\cite{sirinam2018deep}, original model:  \cite{rimmer2017automated})}
   \label{tab:transfer_direction}
   \resizebox{\linewidth}{!}{
   \begin{tabular}{@{}ll@{}}
\toprule
Adversary Strength & Transferability $(\%)$\\
\midrule
$\alpha = 100$ &  $30.65$\\
$\alpha = 500$ &  $85.90$\\
$\alpha = 1000$ & $96.53$\\
\bottomrule
\end{tabular}}
\end{minipage}\hfill
\begin{minipage}{0.3\linewidth}
   \centering
   \caption{Transferability of timing  perturbations (surrogate model: AlexNet, original model:  DeepCorr~\cite{deepcorr})}
   \label{tab:transfer_time}
   \resizebox{\linewidth}{!}{
   \begin{tabular}{@{}ll@{}}
\toprule
Adversary Strength & Transferability $(\%)$\\
\midrule
$\mu = 0, \sigma = 20$ & $46.94$ \\
$\mu = 20, \sigma = 20$ & $76.22$ \\
$\mu = 50, \sigma = 20$ & $88.99$ \\
\bottomrule
\end{tabular}}
\end{minipage}\hfill
\begin{minipage}{0.3\linewidth}
   \centering
   \caption{Transferability of size  perturbations: (surrogate model: AlexNet, original model:  DeepCorr~\cite{deepcorr}) }
   \label{tab:transfer_size}
   \resizebox{1.0\linewidth}{!}{
   \begin{tabular}{@{}ll@{}}
\toprule
Adversary Strength & Transferability $(\%)$\\
\midrule
$N = 10$ & $76.60$ \\
$N = 20$ & $84.54$ \\
$N = 50$ & $91.47$ \\
\bottomrule
\end{tabular}}
\end{minipage}\hfill
\end{table*}

\paragraphb{Our Tailored Defense:}
 We use the adversarial training approach in which the defender uses adversarial perturbations crafted by our attack to make the target model robust against the attacks.  We assume the defender knows the objective function and its parameters. We evaluate our defense when the defender does not know if the attack is targeted or untargeted (for both source and destination). The defender trains the model for one epoch, and then generates blind adversarial perturbations from all possible settings using algorithm~\ref{alg:general_adv}. Then, he extends the training dataset by including all of the adversarial samples generated by the adversary and trains the target model on the augmented train dataset. Algorithm~\ref{alg:defense} in Appendix~\ref{app:def}  sketches our defense algorithm.

\paragraphb{Comparing our defense vs.\ prior defenses:}
 We compare our defense with previous defenses borrowed from the image classification literature.
  Tables~\ref{tab:defense_wf} and~\ref{tab:defense_deep}
  compare the performances of different defenses on DF  and DeepCorr  scenarios, respectively.
 As we see, \textbf{none of the prior defenses for adversarial examples are robust against our blind adversarial attacks}, and in some cases, utilizing them even improves the accuracy of the attack. However, the results show that our tailored defense is more robust than prior defenses. Since the attacker knows the exact attack mechanism, all defense methods cannot perform well when the adversary uses higher strengths in crafting adversarial perturbations. While our defense  is more robust against  blind adversarial attacks,  \emph{it increases the training time of the target model by orders of magnitude} which makes it not scalable for larger models.
 Therefore, designing efficient defenses against blind adversarial perturbations is an important future work.

\section{Transferability}\label{sec:transfer}

An adversarial perturbation scheme is called to be \emph{transferable} if the perturbations it creates for a target model can misclassify other models as well.
A transferable perturbation algorithm is much more practical, as the adversary will not need to have a whitebox access to its target model; instead, the adversary will be able to use 
a \emph{surrogate} (whitebox) model to craft its adversarial perturbations, and then apply them to the \emph{original} blackbox target model.

In this section, we evaluate the transferability of our blind adversarial perturbation technique.
First, we train a surrogate model for our traffic analysis application. Note that, the original and surrogate models do not need to  have the same architecture, but they are trained for the same task (likely with difference classification accuracies).
Next, we create a perturbation generation function $G(z)$ for our surrogate model (as described before). 
We use this $G(z)$ to generate  perturbations, and apply these perturbations on some sample flows. 
Finally,  we feed the resulted perturbed flows as inputs to the \textit{original model} (i.e., the target blackbox model) of the traffic analysis application. 
We measure transferability using a common metric from \cite{Papernot2016TransferabilityIM}: 
we identify the input flows that are correctly classified by both original and surrogate models before applying the blind adversarial perturbation; then, among these samples, we return the ratio of samples  misclassified by the original model over the samples  misclassified by the surrogate model as our transferability metric. 

%

\paragraphb{Direction-based technique}
To evaluate the transferability of our direction-based perturbations, we use the DF system~\cite{sirinam2018deep} as the surrogate model and the WF system of Rimmer et al.~\cite{rimmer2017automated} as the original model.
Note that the model proposed by Rimmer et al.\ uses CNNs, however, it has a completely different structure than DF.
We train both models on DF's dataset~\cite{sirinam2018deep}, and generate blind adversarial perturbations for the surrogate DF model. Then we test the original model using these perturbations. Table~\ref{tab:transfer_direction} shows the transferability of our direction-based attack with different  noise strengths.
As can be seen, our direction-based attack is \emph{highly transferable}.

\paragraphb{Time-based technique}
For the transferability of the time-based attack, we use DeepCorr~\cite{deepcorr} as the original model. 
We use AlexNet~\cite{alexnet} as the  surrogate model, which  has a \emph{completely different architecture}.
We train AlexNet  on the same dataset used by DeepCorr.
Since the main task of AlexNet is image classification, we  modify its hyper-parameters slightly to make it compatible with the DeepCorr dataset.
To calculate  transferability, we fix the false positive rates of both surrogate and original models to the same values (by choosing the right flow correlation thresholds). 
Table~\ref{tab:transfer_time} shows high degrees of transferability for the time-based attack with different blind noise strengths (for a constant false positive rate of $10^{-4}$). 


\paragraphb{Size-based technique}
To evaluate the transferability of the size-based perturbations, we use DeepCorr as the original model and AlexNet as the surrogate model, and calculate transferability as before. Table~\ref{tab:transfer_size} shows the transferability of the size-based technique with different blind noise strengths for a  false positive rate of  $10^{-4}$.

To summarize, we show that  \textbf{blind adversarial perturbations are highly transferable} between different model architectures, allowing them to be used by blackbox adversaries.

\section{Conclusions}

In this paper, we introduced 
blind adversarial perturbations, a mechanism to defeat DNN-based traffic analysis classifiers which works by perturbing the features of live network connections. 
 We presented a systematic approach to generate blind  adversarial perturbations through solving specific optimization problems tailored to traffic analysis applications. 
Our blind adversarial perturbations algorithm is generic and can be applied on various types of traffic classifiers with different network constraints. 

We evaluated our attack against state-of-the-art traffic analysis systems, showing that our attack  outperforms traditional techniques in defeating traffic  analysis.  
We also showed that our blind adversarial perturbations are even transferable between different models and architectures, so they can be applied by blackbox adversaries. 
Finally, we showed that existing defenses against adversarial examples perform poorly against blind adversarial perturbations, therefore we  designed a tailored countermeasure against blind perturbations.

\section*{Acknowledgements}
The project is generously supported by the NSF CAREER grant CNS-1553301 and the NSF grant CNS-1564067. Milad Nasr is supported by a Google PhD Fellowship in Security and Privacy.
\bibliographystyle{plain}
\bibliography{all-bibs}

\appendix

 \section{Adapting Traditional Defenses to Adversarial Examples}\label{sec:prior-defense}

Madry et al.~\cite{madry2017towards} presented a scalable adversarial training approach to increase the robustness of deep learning models to adversarial examples. In each iteration of training, this method generates a set of adversarial examples and uses them in the training phase. Madry et al.'s defense is the most robust defense among the adversarial training based defenses~\cite{carlini2017adversarial}. We cannot use this method as is, since in the image recognition applications, pixels can take real values, while in direction-based traffic analysis methods, features take only two values (-1, +1). Therefore, we  modify this defense to our setting. To generate a set of adversarial examples in the training process, we randomly choose a number of packets and flip their directions from -1 to +1 and vice versa. Similarly, for the packet timings and sizes we enforced all of application constraints for generating the adversarial examples.

From the gradient mask approach, we used the \emph{input gradient regularization} (IGR) technique of Ross and Doshi-Velez~\cite{ross2018improving}. IGR is more robust against adversarial attacks compared to its previous work~\cite{papernot2016distillation}. Their defense trains a model to have smooth input gradients with fewer extreme values which becomes more resistant to adversarial examples. We utilize this approach to train a robust model using DF structure. 
We evaluated the direction-based attack against this defense with parameter $\lambda = 10$.

While the previous defenses train a robust model against adversarial attacks, Cao and Gong~\cite{cao2017mitigating} designed a defense method which does not change the training process. They proposed a \emph{region-based classification} (RC) method which creates a hypercube centered at the input to predict its label. Then, the method samples a set of data points from the crafted hypercube and  uses an existing trained model to produce predicted label for each sampled data point; Finally, it uses majority voting to generate the final class label for the given input. We need to make changes to the region-based classification defense. In contrast to images, we cannot create a hypercube centered at the input by just adding random real values to the packet direction sequences which have values -1, 1. Instead, for each input, we create the hypercube by randomly choosing a number of packets in the sequence and flipping their directions. To apply region-based classification in the test phase of the direction-based method while adding blind perturbations, we randomly choose 125 packets and change their directions to form the hypercube. Similar to Cao and Gong, we call this number as the radius of the hypercube. We choose this value for the radius because 125 is the maximum number of packets we can use to form the hypercube while the accuracy of the region-based method does not go below the accuracy of the original DF model. Using radius of 125 for hypercubes, we apply the region-based classification against our attack. For time and size based methods, we use the strength of the adversary to generate the hypercubes.

\section{Model Parameters}\label{app:params}
Table~\ref{tab:model_param} shows the structure of the adversarial model for each type of perturbation and also the discriminator model.

\begin{table*}[h]
   \centering
\caption{Tuned parameters of the adversarial models and discriminator model}   
   \resizebox{0.8\textwidth}{!}{
   \begin{tabular}{@{}lcccccc@{}}
\toprule
\textbf{Model} & Number of hidden layers & Hidden layer units & Optimizer & Learning rate & Activation \\
\midrule
Direction-based &  $1$ & $[500]$ & \text{Adam} & $10^{-3}$ & \text{ReLu} \\
Time-based &  $1$ & $[500]$ & \text{Adam} & $10^{-3}$ & \text{ReLu} \\
Size-based (ordering)  & $1$ & $[500]$ & \text{Adam} & $10^{-3}$ & \text{ReLu} \\
Size-based (amplitude)  & $1$ & $[500]$ & \text{Adam} & $10^{-3}$ & \text{ReLu}\\
Discriminator  & $2$ & $[1000, 1]$ & \text{Adam} & $10^{-4}$ & \text{ReLu} \\
\bottomrule
\end{tabular}}
\label{tab:model_param}
\vspace{0.2cm}
\end{table*}

\section{The Scheme of Our Blind Adversarial Perturbation Technique}\label{app:scheme}

\begin{figure*}[h]
     \centering
     \includegraphics[width = \linewidth]{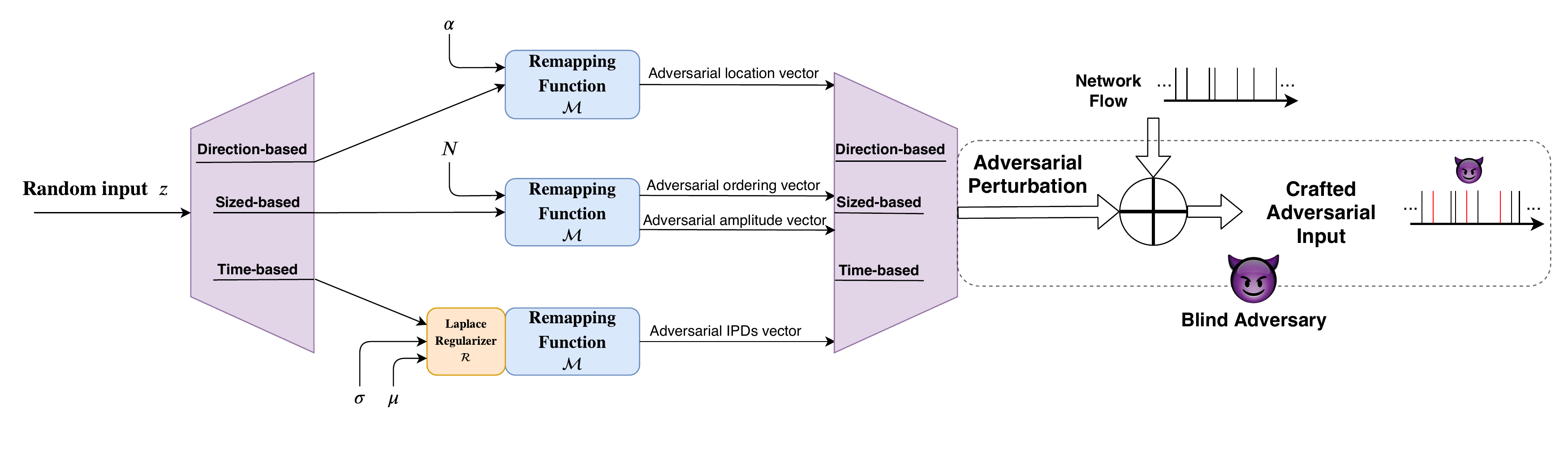}
     \caption{The block diagram of our blind adversarial perturbation technique}
     \label{fig:training}
 \end{figure*}

Figure~\ref{fig:training}  illustrates  the main components of our blind adversarial perturbations algorithm.

%
%

\section{Defense Algorithm}\label{app:def}

Algorithm~\ref{alg:defense} summarizes the adversarial approach to train a robust model against blind adversarial attacks in traffic analysis applications.

\begin{algorithm}[h]
    \caption{Adversarial defense against blind adversarial perturbation}
    \label{alg:defense}
   \begin{algorithmic}
       \STATE Randomly initialize network $N$
       \STATE $\mathcal{L}_{f} \gets $ target model loss function
       \STATE $\mathcal{M} \gets$ domain  remapping function
       \STATE $\mathcal{R} \gets$ domain  regularizations function
       \STATE $G(z) \gets$ initialize the blind adversarial perturbation model parameters ($\theta_G$)
       \STATE $T \gets $ epochs
       \STATE $Z \gets $  [] \MCOMMENT{List of adversarial perturbations}
       \FOR{epoch $t \in \{ 1 \cdots T \}$  }
       \STATE Train the model $N$ for one epoch on training dataset $\mathcal{D}^{tr}$
       \STATE $Z \gets$ generate adversarial perturbations using Algorithm~\ref{alg:general_adv} from all possible targets and focus classeså
       \ENDFOR
       \STATE $\mathcal{D}^{tr}$.extend($\mathcal{D}^{tr} + Z$)
       \RETURN $N$
    \end{algorithmic}
\end{algorithm}

\end{document}